\documentclass[usenatbib]{mn2e}   
\pdfoutput=1
\usepackage[pdftex]{graphicx} 
\usepackage{times}
\usepackage{rotate} 
\usepackage{amssymb} 
\usepackage{rotating}
\usepackage{epstopdf} 

\title[On  the  nature  of  Infrared Excess  Sources]{Infrared  Excess
  Sources: Compton Thick QSOs, low luminosity Seyferts or starbursts?}
\author[Georgakakis  et al.   ]   {A.  Georgakakis$^{1}$\thanks{email:
    age@astro.noa.gr},   M.    Rowan-Robinson$^2$,   K.    Nandra$^2$,
  J.        Digby-North$^2$       \\       \\        {\rm       \LARGE
    P.     G.      P\'erez-Gonz\'alez$^{3,4}$,     G.     Barro$^{3}$}
  \\ \\ $^1$National Observatory of  Athens, V.  Paulou \& I.  Metaxa,
  11532,   Greece\\  $^2$Astrophysics   Group,   Blackett  Laboratory,
  Imperial   College,   Prince   Consort   Rd  ,   London   SW7   2AZ,
  UK\\ $^3$Departamento de  Astrof\'isica, Facultad de CC.  F\'isicas,
  Universidad     Complutense    de     Madrid,     E-28040    Madrid,
  Spain\\ $^4$Department of  Astronomy, Steward Observatory, 933 North
  Cherry Avenue, Tucson, AZ 85721-0065, USA }
\begin{document}
\maketitle

\begin{abstract}  We explore  the  nature of  Infrared Excess  sources
(IRX),  which  are  proposed  as  candidates  for  luminous  [$L_X(\rm
    2-10\,keV)  >  10^{43} \,  erg  \,  s^{-1}$]  Compton Thick  ($\rm
  N_H>2\times  10^{24}  \,   cm^{-2}$)  QSOs  at  $z\approx2$.   Lower
  redshift, $z\approx1$,  analogues of the distant  IRX population are
  identified by firstly  redshifting to $z=2$ the SEDs  of all sources
  with  secure spectroscopic  redshifts in  the AEGIS  (6488)  and the
  GOODS-North (1784) surveys and  then selecting those that qualify as
  IRX sources at that redshift.   A total of 19 galaxies are selected.
  The  mean redshift of  the sample  is $z\approx1$.   We do  not find
  strong evidence for Compton Thick  QSOs in the sample. For 9 sources
  with  X-ray  counterparts, the  X-ray  spectra  are consistent  with
  Compton  Thin AGN.   Only  3  of them  show  tentative evidence  for
  Compton  Thick  obscuration.   The  SEDs  of  the  X-ray  undetected
  population  are consistent  with  starburst activity.   There is  no
  evidence  for a hot  dust component  at the  mid-infrared associated
  with AGN heated dust.  If  the X-ray undetected sources host AGN, an
  upper limit  of $L_X(\rm 2-10\,keV) =  10^{43} \, erg  \, s^{-1}$ is
  estimated for  their intrinsic luminosity.  We propose  that a large
  fraction  of the $z\approx2$  IRX population  are not  Compton Thick
  QSOs but  low luminosity  [$L_X(\rm 2-10\,keV) <  10^{43} \,  erg \,
    s^{-1}$], possibly  Compton Thin, AGN or dusty  starbursts.  It is
  shown that  the decomposition of the AGN  and starburst contribution
  to  the mid-IR  is essential  for  interpreting the  nature of  this
  population, as star-formation may dominate this wavelength regime.
\end{abstract}

\section{Introduction}\label{sec_intro}

The composition of the diffuse  X-ray background remains a problem for
high energy astrophysics.  X-ray  imaging surveys with the Chandra and
XMM-Newton observatories have resolved between  80 to 100\% of the XRB
into discrete  sources below about  6\,keV \citep[e.g.][]{Worsley2005,
  Hickox2006,  Georgakakis2008_sense}.   The  vast majority  of  these
sources are Compton thin AGN  ($\rm N_H \la 10^{24} \rm \,cm^{-2}$) at
a mean redshift $z\approx1$ \citep[e.g.][]{Barger2005, Akylas2006}. At
higher energies however, between 20 and 30\,keV, where the bulk of the
XRB energy  is emitted \citep{Marshall1980}, only a  small fraction of
its  total   intensity  has   been  resolved  into   discreet  sources
\citep{Sazonov2007}.  As  a result the nature of  the populations that
make up the XRB at these energies is still not well known.  Population
synthesis  models use  our knowledge  on the  properties of  the X-ray
sources below  $\approx10$\,keV to make  predictions on the  nature of
the   X-ray    populations   close   to   the   peak    of   the   XRB
\citep[e.g.][]{Gilli2007}.   These models  indicate that  Compton thin
AGN  alone  cannot  account for  the  shape  of  the XRB  spectrum  at
$\approx20-30$\,keV.   An additional  population of  heavily obscured,
Compton thick ($\rm N_H \ga  10^{24} \rm \,cm^{-2}$) AGN is postulated
to reconcile the  discrepancy \citep[e.g.][]{Gilli2007}.  The required
number   density   of   such   sources  is   however,   under   debate
\citep[][]{Treister2009xrb, Draper_Ballantyne2009}.  Unfortunately the
identification of the heavily obscured AGN population predicted by the
models is far from trivial.  The X-ray emission of these sources below
about  10\,keV is  suppressed  by photoelectric  absorption  and as  a
result most  of them  are expected to  lie well below  the sensitivity
limits of the deepest  current X-ray observations.  Although a handful
of Compton  thick AGN  candidates have been  identified in  deep X-ray
surveys  \citep{Tozzi2006,  Georgantopoulos2009},  the  bulk  of  this
population remains to be discovered.

Selection at the mid-IR ($\rm  3-30\,\mu m$) is proposed as a powerful
tool  for finding heavily  obscured X-ray  faint AGN.   The UV/optical
photons emitted by the central engine are absorbed by the gas and dust
clouds and appear as thermal radiation with a broad bump in the mid-IR
\citep[$\nu f_{\nu}$ units; e.g.  ][]{Elvis1994, Prieto2009}.  Diverse
selection methods  have been developed  to identify this  AGN spectral
signature in the  mid-infrared.  \cite{Lacy2004}, \cite{Stern2005} and
\cite{Hatziminaoglou2005}  propose  simple colour  cuts  based on  the
mid-IR colours of luminous  high redshift QSOs and/or type-2 Seyferts.
\cite{Polletta2006} and \cite{Rowan-Robinson2009} fit templates to the
broad-band     photometry      from     UV     to      the     far-IR.
\cite{Alonso-Herrero2006}  and \cite{Donley2007}  select  sources with
power-law  Spectral Energy  Distributions (SEDs)  in the  mid-IR.  The
methods  above  have merits  and  shortcomings.   Selection by  mid-IR
colour  for   example,  is  simple  but   suffers  contamination  from
star-forming  galaxies  if  applied  to  the  deepest  mid-IR  samples
currently available  \citep{Georgantopoulos2008, Donley2008}. Template
fits  are powerful  but require  high quality  photometry over  a wide
wavelength  baseline  for   meaningful  constraints.   Power-law  SEDs
provide  the most  clean samples  of  infrared selected  AGN, but  are
sensitive only to the most luminous and hence, rare sources.

\begin{figure}
\begin{center}
\includegraphics[height=0.9\columnwidth]{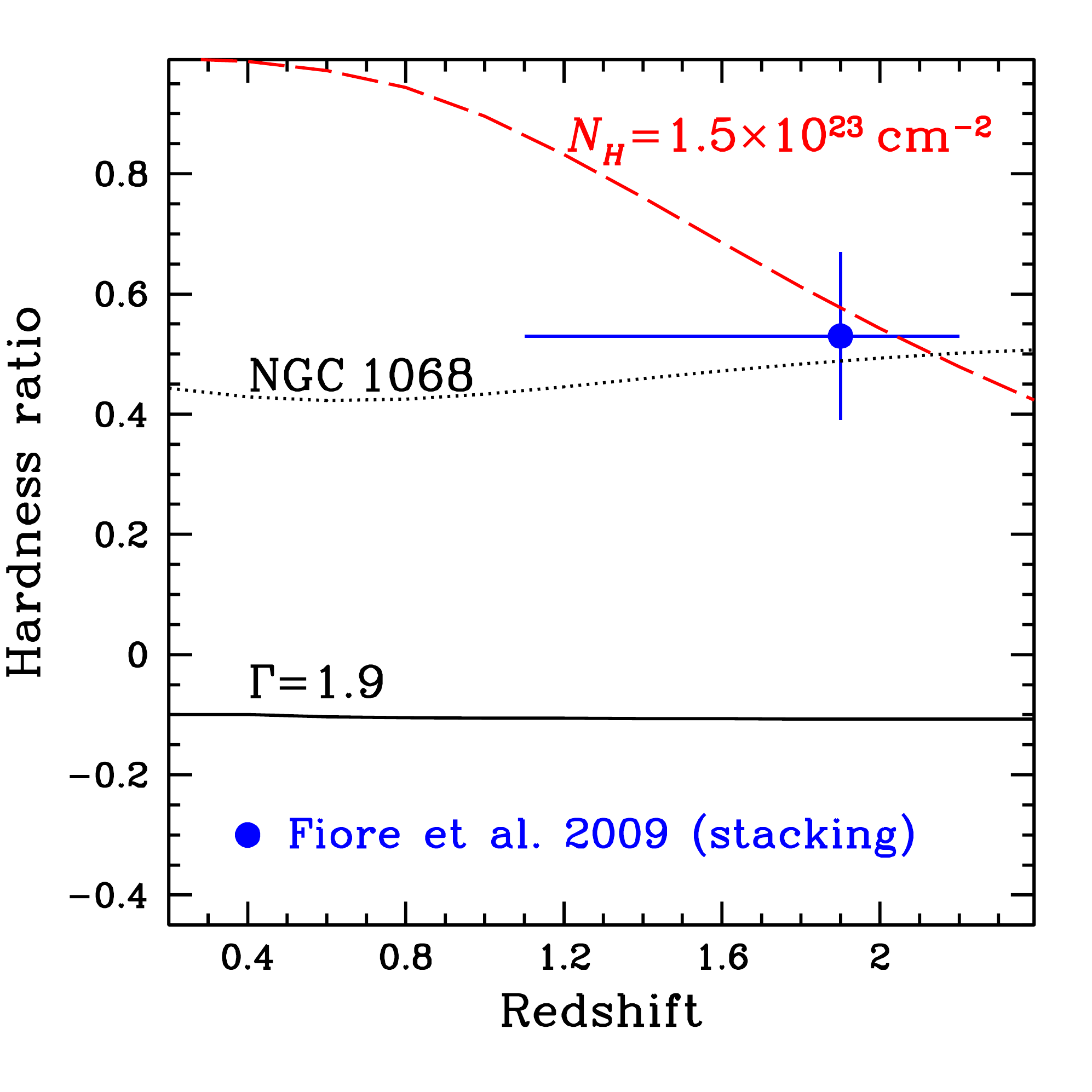}
\end{center}
\caption{Hardness  ratio  against  redshift.   The hardness  ratio  is
  defined as HR=(H-S)/(H+S), where H and S are the counts in the 1.5-6
  and 0.3-1.5\,keV  energy intervals respectively.  The  (blue) dot is
  the mean  hardness ratio of  X-ray undetected sources in  the COSMOS
  survey (Fiore  et al. 2009) estimated using  stacking analysis.  The
  vertical errorbar is the $1\sigma$  uncertainty of the HR, while the
  horizontal  errorbar  shows  the  redshift  range of  the  Fiore  et
  al.  (2009)  IRX population.   The  dotted  line  (black) shows  the
  expected  HR  of the  local  Compton  Thick  AGN NGC\,1068  Matt  et
  al.  (1999). The  (red)  dashed curve  corresponds  to the  HR of  a
  power-law X-ray  spectrum with  photon index $\Gamma=1.9$,  which is
  absorbed  by a column  density of  $\rm N_H=  1.5 \times  10^{23} \,
  cm^{-2}$. The  continuous (black) line  corresponds to the HR  of an
  unobscured  AGN, i.e a  power-law X-ray  spectrum with  photon index
  $\Gamma=1.9$.  }\label{fig_hr}
\end{figure}

\begin{figure}
\begin{center}
\rotatebox{0}{\includegraphics[height=0.9\columnwidth]{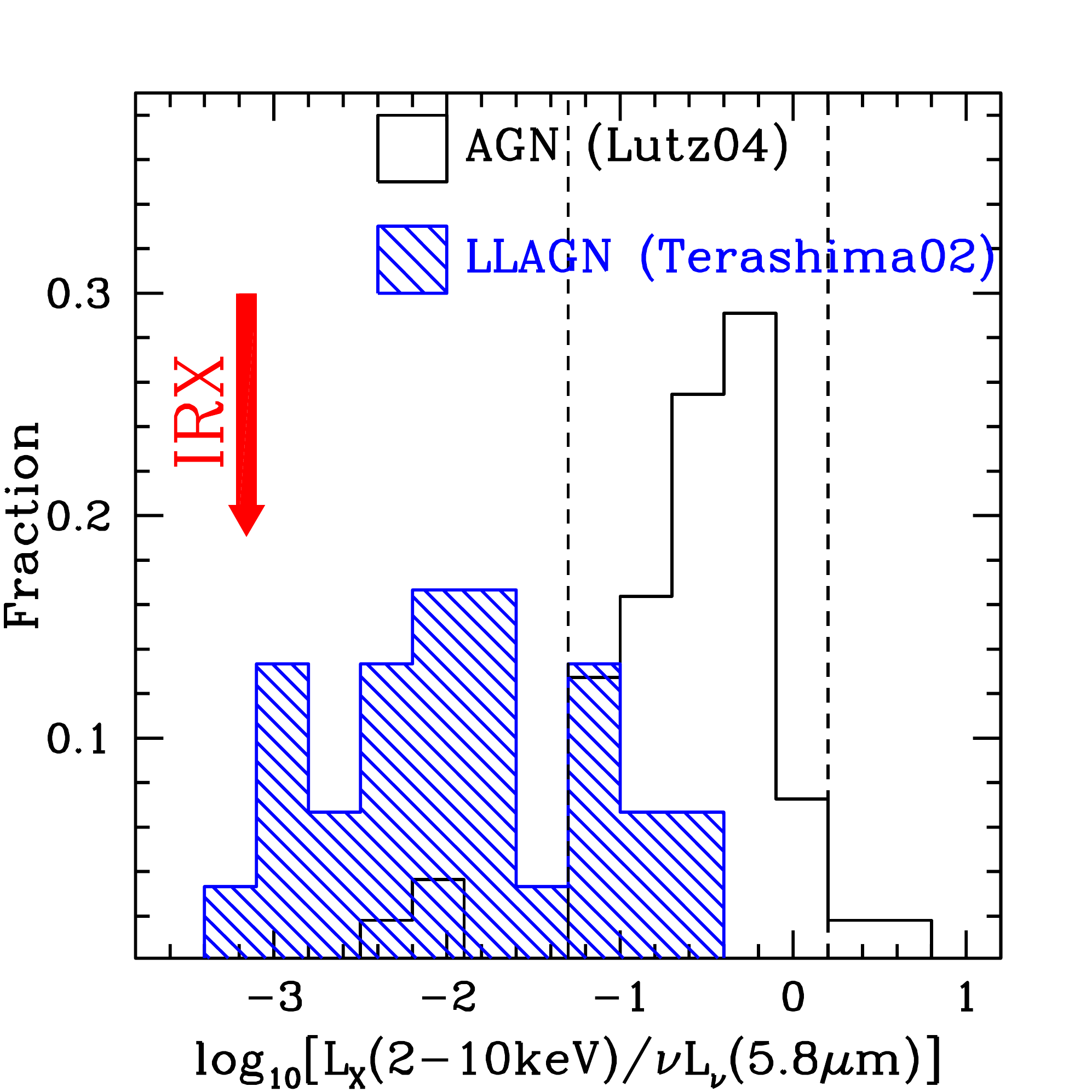}}
\end{center}
\caption{Distribution of X-ray--to--mid-IR luminosity ratio, $L_X({\rm
    2-10\,keV})/  \nu L_{\nu}(\rm  5.8\mu m)$.   The  (black) unshaded
  histogram  is for  AGN from  Lutz et  al. (2004).   This  sample was
  selected  to include  only AGN  that  dominate in  the mid-IR.   The
  vertical dashed  lines show the dispersion  of the X-ray--to--mid-IR
  luminosity  ratio for the  same sample.   The low-luminosity  AGN of
  Terashima et al.  (2002) are  shown with the blue hatched histogram.
  The  mean  $L_X({\rm  2-10\,keV})/  \nu L_{\nu}(5.8\mu  m)$  of  IRX
  sources is plotted with the (red) arrow.  This estimate is for X-ray
  undetected  IRX sources  in  the CDF-North  (Georgantopoulos et  al.
  2008). The  mean X-ray luminosity  of this population  is determined
  via stacking analysis and $\nu L_{\nu}(\rm 5.8\mu m)$ is the average
  mid-IR luminosity  of the  sample.  Both luminosities  are estimated
  assuming $z=2$ for  the IRX sources.  The $\rm  5.8\mu m$ luminosity
  of the Terashima et al.  (2002) sources is determined using the IRAS
  $\rm 12\,\mu  m$ flux density  and adopting an average  flux density
  ratio $f_{\rm 12\,\mu m}/f_{\rm 5.8\,\mu m}=0.9$, estimated from the
  subsample of  these sources with $6\mu m$  flux density measurements
  from  the  literature.   The  2-10\,keV X-ray  luminosities  of  the
  Terashima et  al.  (2002) AGN are corrected  for obscuration, except
  for a total of 6 sources  in that sample, which show a strong Fe\,Ka
  6.4\,keV line in their X-ray spectra with equivalent width $\rm EW >
  900   \,   eV$  and   are   therefore   Compton  Thick   candidates.
}\label{fig_hist}
\end{figure}

A  much  promising  method  that   is  believed  to  be  efficient  in
identifying heavily obscured, possibly  Compton Thick, AGN is based on
the  selection of  sources that  are faint  at optical  and  bright at
mid-IR wavelengths.  In its simplest version this method applies a cut
in the $\rm 24\,\mu m$ over $R$-band flux density ratio, $f_{\rm 24\mu
  m}  /  f_{\rm  R}  >   1000$  to  identify  Dust  Obscured  Galaxies
\citep[DOGs;][]{Dey2008} at a  mean redshift $z\approx2$.  In addition
to the limit above \cite{Fiore2008, Fiore2009} also use the colour cut
$R-K>4.5$ to select Infrared-Excess  sources (IRXs). This colour selection is
motivated by  the observational result  that redder sources  include a
higher fraction of obscured AGN \citep[e.g.][]{Brusa2005}. A different
approach  has been  adopted  by \cite{Daddi2007}.   They select  $BzK$
sources  that show excess  mid-IR emission  relative to  that expected
based  on   the  rates  of   star  formation  measured   from  shorter
wavelengths.   Despite differences  in the  adopted criteria,  all the
studies  above   identify  a  population  of   galaxies  with  similar
properties  in terms  of  average redshift  ($z\approx2$), mean  X-ray
properties and mid-IR luminosities.   These sources are believed to be
massive  galaxies  \citep[$M_{star}   \approx  10^{10}  -  10^{11}  \,
  M_{\odot}$][]{Treister2009,    Bussmann2009_hst}   that   experience
intense  bursts of  star-formation and  rapid supermassive  black hole
growth, possibly triggered by mergers \citep[e.g.][]{Bussmann2009_hst,
  Narayanan2009}.  The  apparently brighter subset  of this population
($\rm   S_{24}>300\,  \mu   Jy$)  are   proposed  as   descendants  of
Submillimeter  Galaxies  on  their  way to  becoming  unobscured  QSOs
\citep[e.g.][]{Bussmann2009_hst,   Bussmann2009_sed,   Narayanan2009},
which will eventually  evolve    into   present-day    $4\,L^{*}$   galaxies
\citep{Dey2008, Brodwin2008}.  Although these sources  are undoubtedly
important for  understanding the co-evolution of galaxies  and SBHs at
high redshift,  in this  paper we focus  on their significance  to XRB
studies by putting into test claims that they include a large fraction
of  Compton  Thick AGN.   The  two  key  properties of  these  sources
(hereafter  referred to  as  IRX sources),  which  are interpreted  as
evidence for Compton Thick AGN  are (i) their hard mean X-ray spectrum
and (ii) their faintness at  X-ray wavelengths relative to the mid-IR.
The first point is demonstrated in Figure \ref{fig_hr} which shows the
that the  mean hardness  ratio of the  IRX sources is  consistent with
that of  the local Compton Thick AGN  NGC\,1068 \citep{Matt1999}. With
respect to the second point, the majority of the IRX sources \citep[80
  per  cent;][]{Georgantopoulos2008} are not  detected in  the deepest
X-ray surveys  available and their  mean X-ray properties can  only be
studied through  stacking analysis.  Figure  \ref{fig_hist}, shows the
average X-ray to mid-IR  luminosity ratio, $L_X({\rm 2-10\,keV}) / \nu
L_{\nu}  {\rm 5.8\mu  m}$,  of the  X-ray  undetected IRX  population.
Relative to local AGN from the sample of \cite{Lutz2004} these sources
appear  underluminous in the  mid-IR by  2-3\,dex.  Assuming  a narrow
range  of $L_X({\rm  2-10\,keV}) /  \nu  L_{\nu} {\rm  5.8\mu m}$  for
typical  AGN  (e.g.  Lutz  et  al.  2004),  the  position  of the  IRX
population in Figure \ref{fig_hist} is consistent with a Compton Thick
obscuring screen that suppresses  the observed X-ray emission relative
to the mid-IR.   Under the assumption that the  mid-IR luminosity is a
good  proxy of  the AGN  power, one  can estimate  intrinsic 2-10\,keV
luminosities  for the IRX  sources in  excess of  $10^{43} \rm  erg \,
s^{-1}$ \citep{Daddi2007,Fiore2008,Fiore2009, Treister2009}.  AGN with
intrinsic luminosities  above this limit are hereafter  referred to as
QSOs.  IRX sources are therefore, excellent candidates for the Compton
Thick QSOs  needed by population  synthesis models to explain  the XRB
spectrum at 20-30\,keV. Simulations indeed show that the bulk of these
sources (80  per cent, Fiore  et al. 2008;  95 per cent, Fiore  et al.
2009) are likely to be Compton Thick QSOs, while their estimated space
densities are consistent with the predictions of the XRB models.

Although there is no doubt that the IRX population includes a fraction
of heavily obscured  AGN \citep{Georgantopoulos2009}, the evidence for
Compton Thick sources is far  from conclusive.  The two key properties
of  the IRX  sources, hardness  ratio and  X-ray to  mid-IR luminosity
ratio, are also consistent with  Compton thin AGN of lower luminosity,
hereafter defined as $L_X(\rm 2-10\,keV) < 10^{43} \rm erg \, s^{-1}$.
Figure \ref{fig_hr} for example, shows that the mean hardness ratio of
this population could be due to moderate obscuring column densities of
few  times $\rm  10^{23}\,cm^{-2}$.   Moreover, Figure  \ref{fig_hist}
demonstrates that lower luminosity  Compton Thin AGN (Terashima et al.
2002) have very  low X-ray to mid-IR luminosity  ratios, i.e.  similar
to those observed for IRX sources.  This is not surprising as both AGN
and star-formation  contribute to the  mid-IR, thereby resulting  to a
broad   $L_X({\rm  2-10\,keV})   /   \nu  L_{\nu}   {\rm  5.8\mu   m}$
distribution.  One has to isolate the AGN component at the mid-IR part
of  the  SED  to  get   a  tight  correlation  with  X-ray  luminosity
\citep[e.g.     ][]{Prieto2009}.     \cite{Lutz2004}   for    example,
accomplished  that  by selecting  {\it  only}  sources  where the  AGN
dominates  in the  mid-IR,  i.e.  those  {\it  without} a  significant
starburst component  relative to  the AGN. For  the high  redshift IRX
sources it is  not clear what fraction of the  mid-IR emission is from
AGN heated dust.  Only under the strong assumption that the {\it bulk}
the mid-IR  luminosity is  associated with reprocessed  radiation from
accretion on  the central  SBH, can one  infer that these  sources are
Compton  Thick QSOs.   A number  of studies  on the  IRX  sources have
already suggested that this population may include a large fraction of
lower   luminosity  Compton   thin  AGN   or  even   dusty  starbursts
\citep[e.g.][]{Georgantopoulos2008, Donley2008, Pope2008, Murphy2009}.

Elucidating  the nature of  IRX sources,  Compton Thick  QSOs, Compton
Thin lower luminosity AGN or starbursts, requires very deep X-ray data
and/or AGN/starburst decomposition  at the mid-IR.  Unfortunately, IRX
samples at $z\approx2$ are too  faint at almost any wavelength, expect
the mid-IR, for  such a detailed study. In this  paper we address this
issue  by  selecting lower  redshift  ($z\approx1$)  analogues of  the
distant ($z\approx2$)  IRX population. The advantage  of this approach
is that the selected sources are apparently bright and therefore their
SEDs  can be  constrained over  a wide  wavelength baseline,  from the
far-UV to the far-IR. The sources  are selected in fields with some of
the deepest  X-ray observations  available (Chandra Deep  Field North,
All  wavelength  Extended Groth  strip  International Survey:  AEGIS),
thereby  allowing  study  of   their  X-ray  properties  via  spectral
analysis. An  additional advantage  of our strategy  is that  there is
only  about 2.5\,Gyr  difference in  the age  of the  Universe between
$z=1$ and  $z=2$ ($\rm  H_0 = 70  \, km  \, s^{-1} \,  Mpc^{-1}$, $\rm
\Omega_{M}  =  0.3$, $\rm  \Omega_{\Lambda}  =  0.7$).  Therefore,  by
selecting  galaxies at  $z\approx1$,  it is  more  likely to  identify
systems  that  are  physically   similar  to  the  IRX  population  at
$\approx2$.   In this  respect it  is  interesting that  in the  local
Universe there are no known analogues of the IRX sources, which can be
used   to   study   in   detail   the  nature   of   this   population
\citep[e.g. ][]{Dey2008}.

Throughout  this paper  we adopt  $\rm H_0  = 70  \, km  \,  s^{-1} \,
Mpc^{-1}$, $\rm \Omega_{M} = 0.3$ and $\rm \Omega_{\Lambda} = 0.7$.

\section{Data}\label{sec_data}

The UV  to far-IR  data used  in this study  are from  the Universidad
Complutense      de     Madrid     (UCM)      Cosmological     Surveys
Database\footnote{http://guaix.fis.ucm.es/$\sim$pgperez},         which
combines all the  photometric and spectroscopic observations available
in prime  cosmological survey fields.  Details on  the data reduction,
band  merging   and  photometry  are   presented  by  \cite{Perez2005,
  Perez2008}.  The fields  of choice for this paper  are the AEGIS and
GOODS-North.   In  both fields  the  IRAC  $\rm  3.6\,\mu m$  selected
catalogues are  used, which reach  limiting magnitudes (AB  system) of
[3.6]=25.5  and 24.5\,mag  respectively.  A  total of  80237  and 9061
sources are detected in the  AEGIS and GOODS-North respectively to the
limits above.

The multiwavelength data in AEGIS  include (i) photometry in the GALEX
far- and  near-ultraviolet wavebands (FUV and  NUV respectively; Davis
et al.   2007), (ii) $BRI$ observations  from the Canada-France-Hawaii
Telescope  CFHT12K  mosaic   camera  \citep{Coil2004},  (iii)  $ugriz$
imaging  from  the  T0003  data release  of  the  Canada-France-Hawaii
Telescope  Legacy Survey  \citep[CFHTLS;][]{Ilbert2006},  (iv) Spitzer
IRAC  observations   at  3.6,   4.5,  5.8  and   $8.9\,  \rm   \mu  m$
\citep{Barmby2008} obtained  as part of the  Guaranteed Time Observing
(GTO) program and  (v) Spitzer MIPS photometry at  24 and $\rm 70\,\mu
m$ by coadding GTO data with  the 2nd data release of the Far-Infrared
Deep Extragalactic Legacy Survey (FIDEL). Spectroscopy in the AEGIS is
primarily  from the  DEEP2 survey  \citep{Davis2003, Davis2007}  and a
number of  smaller spectroscopic  programs that targeted  the original
Groth Strip \citep[][and references therein]{Weiner2005}.

The photometry in the GOODS-North includes (i) GALEX FUV and NUV data,
(ii)  deep  $UBVRIz$  ground  based  imaging  presented  by  Capak  et
al. (2004), (iii)  Spitzer IRAC (3.6, 4.5, 5.8 and  $8.9\, \rm \mu m$)
and MIPS $\rm 24\,\mu m$ observations obtained as part of the GOODS
program, MIPS  $\rm 70\,\mu m$ data from  FIDEL.  Optical spectroscopy
in the GOODS-North is available from 
either  programmes that  specifically target  the X-ray  population in
this  field \citep[e.g.][]{Barger2003,  Barger2005, Cowie2003}  or the
Keck Treasury Redshift Survey \citep[TKRS;][]{Wirth2004}.

The X-ray  observations of the Extended Groth  Strip (AEGIS-X) consist
of 8 ACIS-I (Advanced CCD Imaging Spectrometer) pointings, each with a
total integration  time of about 200\,ks  split in at  least 3 shorter
exposures  obtained  at different  epochs.   The  GOODS-North lies  in
central  and most  sensitive region  of the  Chandra Deep  Field North
(CDF-N) survey,  which consists of 20  individual ACIS-I observations,
which sum up  to a total exposure time of 2Ms.   The CDF-N and AEGIS-X
data reduction, source detection and flux estimation have been carried
out  in   a  homogeneous  way  using  the   methodology  described  by
\cite{Laird2009}\footnote{data               available              at
  http://astro.imperial.ac.uk/research/xray/}.   Sources  with Poisson
probabilities $<4\times10^{-6}$  are detected in 4  energy bands, full
(0.5-7\,keV),  soft  (0.5-2\,keV),   hard  (2-7\,keV)  and  ultra-hard
(5-7\,keV).   The  count  rates  in  the above  energy  intervals  are
converted  to fluxes  in the  standard bands  0.5-10, 0.5-2,  2-10 and
5-10\,keV, respectively, assuming a power-law X-ray spectrum with $\rm
\Gamma=1.4$   absorbed  by  the   Galactic  hydrogen   column  density
appropriate  for   each  field.    The  2-10\,keV  fluxes   where  the
sensitivity curves drop at half their maxima are $\rm 2\times10^{-14}$
and $2\times10^{-16} \rm \, erg  \, s^{-1} \, cm^{-2}$ for the AEGIS-X
and CDF-N surveys respectively. The identification of the X-ray sample
with the IRAC $\rm 3.6\,\mu m$  sources has been carried out using the
The Likelihood Ratio method \citep[LR;][]{Sutherland_and_Saunders1992,
  Laird2009, Georgakakis2009}.  In the AEGIS, of the 918 X-ray sources
that overlap  with the Spitzer IRAC  survey region, 867  (93 per cent)
have  an IRAC counterpart  with likelihood  $LR>0.5$ and  an estimated
contamination rate of 1.5 per cent. In the CDF-North 335 X-ray sources
overlap with  the GOODS-North  IRAC data, of  which 301 (90  per cent)
have counterparts  with $LR>0.5$.  The expected  contamination rate is
about 1.4 per cent.

\begin{table*}
\caption{Sample}\label{tab_sample}
\begin{tabular}{l c c c c c  ccccc}
\hline 

ID & RA & Dec & $z$ & $f_{\rm 24\mu m}/f_{R}$ & $R-[3.6]$ & $\log L_{sb}$ & $\log L_{tor}$ & type & $A_V$ & $\rm \log SFR$\\ &

(J2000) & (J2000)&  & ($z=2$) & ($z=2$) &  ($L_\odot$) & ($L_\odot$) &
&(mag)& ($\rm M_\odot/yr$)\\ 

\hline irx-1 &  14 16 17.37 & $+$52  12 38.25 & 0.6826 & 2172  & 5.7 &
11.37 & -- & E & -- & 1.67\\ 

irx-2 & 14  16 32.79 & $+$52 19 01.94  & 1.0284 & 984 &  4.1 & 11.62 &
11.62 & QSO & 0.8 & 1.92 \\ 

irx-3 & 14 16 21.80 & $+$52 20 06.35 & 0.7602 & 1587 & 5.8 & 11.39
& -- & E & -- & 1.69 \\ 

irx-4 & 14 16 18.74 & $+$52 23 19.24 & 0.8360 & 1188 & 6.1 & 11.40 & -- & E & -- & 1.70\\

irx-5 & 14 16  50.49 & $+$52 16 35.11 & 0.6829 & 1678  & 6.0 & 10.87 &
10.57 & E & -- & 1.17\\ 

irx-6  & 14 16 51.00 & $+$52 23 17.68 & 1.0246
& 1103 & 5.8 & 11.42 & -- & E & -- & 1.72\\ 

irx-7 & 14  17 23.10 &
$+$52  27 49.57  & 0.9024  &  1891 &  6.1 &  11.41  & --  & E  & --  &
1.71\\ 

irx-8 &  14 17 41.87 & $+$52  28 23.42 & 1.1482 & 1753  & 5.3 &
12.34 & 12.26 & E & -- & 2.64\\ 

irx-9 & 14 17 45.94 & $+$52 30 32.57 &
0.9853 & 1417 & 5.2 & 11.62 & 11.12 & E & -- & 1.92\\ 

irx-10& 14 18
21.86 & $+$52 38 42.08 & 0.7197 & 1365 & 5.7 & 11.23 & & E & -- & 1.53
\\ 

irx-11& 14 19 38.59 & $+$52 55  53.69 & 0.8976 & 1661 & 4.6 & 11.31
& -- & Sab & -- &1.61 \\ 

irx-12& 14 20 38.87 & $+$53 08 16.21 & 0.6873
& 1531 & 5.7 & 11.07 & 10.77 & E & -- & 1.37 \\

irx-13& 14 21 16.46  & $+$53 05 27.55 & 0.7433 & 1282  & 6.0 & 11.07 &
-- & E & -- & 1.37\\ 

irx-14& 14 22 38.98 & $+$53  24 14.80 & 1.2873 &
2984 & 4.6 & 12.49 & -- & E & -- & 2.79\\ 

& & & & & \\ 

irx-15 & 12 36
29.13 & $+$62 10 45.95  & 1.0130 & 1055 & 5.6 & 11.91 &  -- & E & -- &
1.89\\ 

irx-16 & 12  36 56.01 & $+$62 08 07.86 & 0.7920  & 1136 & 5.7 &
11.78 & -- & E & -- & 2.08\\ 

irx-17 & 12 36 46.61  & $+$62 10 48.44 &
0.9399 & 912 & 5.1 & 11.41 & -- & Sab & 0.35 & 1.39\\ 

irx-18 & 12 37
16.65 & $+$62 17  33.56 & 1.1460 & 1112 & 4.5 & 11.48  & 11.23 & Sab &
0.2 & 1.78\\ 

irx-19 & 12 37 48.67  & $+$62 13 12.84 & 0.9110 & 1641 &
5.5 & -- & 11.75 & QSO & 1.50 & --\\ 

\hline

\end{tabular} 
\begin{list}{}{}
\item  The  columns  are:   (1):  Source  identification;  (2):  Right
  Ascension  (J2000)   of  the  IRAC  $\rm  3.6\,mu   m$  source;  (3)
  Declination  (J2000)  of  the  IRAC  $\rm  3.6\,mu  m$  source;  (4)
  spectroscopic  redshift; (5) observed  $f_{\rm 24\mu  m}/f_{R}$ flux
  ratio if the source were at $z=2$; (6) observed $R-[3.6]$ colour (AB
  system) if  the source were at  $z=2$; (7) log of  the IR luminosity
  ($3-1000\mu m$)  of the starburst component, estimated  from the SED
  spectral fits. A dash indicates  that this component is not required
  to fit the data; (8) log of the IR luminosity ($3-1000\mu m$) of the
  AGN torus  component, estimated from  the SED spectral fits.  A dash
  indicates that this  component is not required to  fit the data; (9)
  best-fit optical template in the optical; (10) dust extinction; (11)
  log of the  star-formation rate estimated from the  SED fits. A dash
  indicates no dust reddening.
\end{list}
\end{table*}

\begin{table*}
\scriptsize
\caption{     Multiwavelength     SEDs     of    the     AEGIS     IRX
  sources}\label{tab_photo_aegis}
\scriptsize
\begin{tabular}{l   cc cc cc cc cc  cccc cc}
\hline ID  & $u$ &  $B$ & $g$  & $R$ &  $r$ & $I$ &  $i$ & $z$  & $\rm
3.6\mu m$  & $\rm 4.5\mu m$  & $\rm 5.8\mu  m$& $\rm 8.0\mu m$  & $\rm
24\mu  m$  &  $\rm 70  \mu  m$  \\  \hline  irx-1 &  $-0.92\pm0.12$  &
$-0.81\pm0.10$  &  $-0.33\pm0.04$ &  $0.45\pm0.01$  & $0.31\pm0.01$  &
$0.93\pm0.01$  &  $0.80\pm0.01$  &  $0.97\pm0.01$  &  $1.63\pm0.01$  &
$1.43\pm0.01$ & $1.35\pm0.01$ & $1.29\pm0.02$ & $-$& $3.45\pm0.05$\\

irx-2 & $0.10\pm0.03$ &  $0.04\pm0.01$ & $0.23\pm0.01$ & $0.43\pm0.01$
&  $0.48\pm0.01$ &  $0.76\pm0.01$  & $0.82\pm0.01$  & $0.95\pm0.01$  &
$1.88\pm0.01$  &  $2.06\pm0.01$  &  $2.22\pm0.01$  &  $2.68\pm0.01$  &
$3.23\pm0.01$& $3.70\pm0.03$\\

irx-3   &   $-0.90\pm0.12$  &   $-0.70\pm0.10$   &  $-0.48\pm0.04$   &
$0.27\pm0.01$  &  $0.08\pm0.02$  &  $0.74\pm0.01$  &  $0.56\pm0.01$  &
$0.71\pm0.01$  &  $1.80\pm0.01$  &  $1.64\pm0.01$  &  $1.61\pm0.01$  &
$1.66\pm0.01$ & $2.73\pm0.01$& $3.50\pm0.04$\\

irx-4   &   $-0.82\pm0.22$  &   $-0.46\pm0.10$   &  $-0.21\pm0.03$   &
$0.41\pm0.01$  &  $0.32\pm0.01$  &  $0.98\pm0.01$  &  $0.85\pm0.01$  &
$1.09\pm0.01$  &  $2.09\pm0.01$  &  $1.95\pm0.01$  &  $1.83\pm0.01$  &
$1.92\pm0.01$ & $2.76\pm0.01$& $3.33\pm0.04$\\

irx-5   &   $-0.86\pm0.14$   &   $-0.36\pm0.04$  &   $0.03\pm0.02$   &
$0.75\pm0.01$  &  $0.64\pm0.01$  &  $1.16\pm0.01$  &  $1.09\pm0.01$  &
$1.23\pm0.01$  &  $1.90\pm0.01$  &  $1.85\pm0.01$  &  $1.89\pm0.01$  &
$2.12\pm0.01$ & $2.76\pm0.01$& $3.36\pm0.04$\\

irx-6   &   $-0.83\pm0.10$  &   $-0.83\pm0.16$   &  $-0.41\pm0.04$   &
$-0.04\pm0.02$  &  $-0.01\pm0.02$ &  $0.54\pm0.01$  & $0.40\pm0.01$  &
$0.65\pm0.01$  &  $1.72\pm0.01$  &  $1.64\pm0.01$  &  $1.44\pm0.01$  &
$1.55\pm0.01$ & $2.45\pm0.01$& $-$\\

irx-7   &   $-0.84\pm0.10$  &   $-0.98\pm0.20$   &  $-0.38\pm0.03$   &
$0.15\pm0.03$  &  $0.00\pm0.02$  &  $0.65\pm0.01$  &  $0.53\pm0.01$  &
$0.74\pm0.01$  &  $1.75\pm0.04$  &  $1.61\pm0.04$  &  $1.48\pm0.04$  &
$1.53\pm0.04$ & $2.65\pm0.01$& $-$\\

irx-8 & $0.26\pm0.02$ &  $0.11\pm0.01$ & $0.41\pm0.01$ & $0.93\pm0.01$
&  $0.84\pm0.01$ &  $1.39\pm0.01$  & $1.27\pm0.01$  & $1.49\pm0.01$  &
$2.69\pm0.01$  &  $2.89\pm0.01$  &  $3.05\pm0.01$  &  $3.30\pm0.01$  &
$3.78\pm0.01$& $4.14\pm0.01$\\

irx-9   &   $-0.71\pm0.14$  &   $-0.35\pm0.04$   &  $-0.25\pm0.03$   &
$0.38\pm0.02$  &  $0.32\pm0.01$  &  $0.98\pm0.01$  &  $0.85\pm0.01$  &
$1.13\pm0.01$  &  $1.92\pm0.01$  &  $1.92\pm0.01$  &  $2.01\pm0.01$  &
$2.33\pm0.01$ & $2.98\pm0.01$& $3.59\pm0.04$\\

irx-10   &  $-1.05\pm0.16$   &  $-0.60\pm0.08$   &   $-0.44\pm0.04$  &
$0.23\pm0.02$  &  $0.02\pm0.02$  &  $0.59\pm0.01$  &  $0.45\pm0.01$  &
$0.63\pm0.01$  &  $1.68\pm0.01$  &  $1.51\pm0.01$  &  $1.55\pm0.01$  &
$1.55\pm0.01$ & $2.60\pm0.01$& $3.53\pm0.04$\\

irx-11   &  $-0.86\pm0.13$   &  $-0.70\pm0.11$   &   $-0.33\pm0.03$  &
$0.12\pm0.01$  &  $0.07\pm0.02$  &  $0.56\pm0.01$  &  $0.45\pm0.01$  &
$0.63\pm0.01$  &  $1.31\pm0.04$  &  $1.10\pm0.04$  &  $0.94\pm0.04$  &
$1.01\pm0.04$ & $-$& $3.38\pm0.05$\\

irx-12   &  $-0.58\pm0.06$   &  $-0.46\pm0.08$   &   $-0.06\pm0.02$  &
$0.62\pm0.01$  &  $0.49\pm0.01$  &  $1.07\pm0.01$  &  $0.92\pm0.01$  &
$1.09\pm0.01$  &  $2.03\pm0.01$  &  $2.03\pm0.01$  &  $2.09\pm0.01$  &
$2.35\pm0.01$ & $2.88\pm0.01$& $3.67\pm0.02$\\

irx-13   &  $-1.39\pm0.45$   &  $-0.58\pm0.16$   &   $-0.55\pm0.04$  &
$0.38\pm0.01$  &  $0.06\pm0.02$  &  $0.86\pm0.01$  &  $0.57\pm0.01$  &
$0.76\pm0.01$  &  $1.71\pm0.01$  &  $1.54\pm0.01$  &  $1.51\pm0.01$  &
$1.51\pm0.01$  &   $2.46\pm0.01$&  $3.34\pm0.05$\\  irx-14   &  $-$  &
$-0.43\pm0.06$ & $-$ & $0.08\pm0.02$ & $-$ & $0.51\pm0.04$ & $-$ & $-$
& $1.75\pm0.04$ & $1.80\pm0.04$ & $1.65\pm0.04$ & $1.89\pm0.04$ & $-$&
$4.00\pm0.01$\\ \hline
\end{tabular} 
\begin{list}{}{}
\item None of the sources are detected in the GALEX FUV band. Source
  irx-8 has a GALEX NUV log flux density of $0.10\pm0.03$ ($\rm \mu
  Jy$ units). The columns are: (1): Source identification; (2): log of
  CFHTLS $u$-band flux density in units of $\rm \mu Jy$; (3): log of
  DEEP2 $B$-band flux density in units of $\rm \mu Jy$; (4): log of
  CFHTLS $g$-band flux density in units of $\rm \mu Jy$; (5): log of
  DEEP2 $R$-band flux density in units of $\rm \mu Jy$; (6): log of
  CFHTLS $r$-band flux density in units of $\rm \mu Jy$; (7): log of
  DEEP2 $I$-band flux density in units of $\rm \mu Jy$; (8): log of
  CFHTLS $i$-band flux density in units of $\rm \mu Jy$; (9): log of
  CFHTLS $z$-band flux density in units of $\rm \mu Jy$; (10): log of
  IRAC $\rm 3.6\mu m$-band flux density in units of $\rm \mu Jy$;
  (11): log of IRAC $\rm 4.5\mu m$-band flux density in units of $\rm
  \mu Jy$; (12): log of IRAC $\rm 5.8\mu m$-band flux density in units
  of $\rm \mu Jy$; (13): log of IRAC $\rm 8.9\mu m$-band flux density
  in units of $\rm \mu Jy$; (14): log of MIPS $\rm 24\mu m$-band flux
  density in units of $\rm \mu Jy$; (15): log of MIPS $\rm 70\mu
  m$-band flux density in units of $\rm \mu Jy$.
\end{list}
\end{table*}

\begin{table*}
\caption{Multiwavelength     SEDs     of     the     CDF-North     IRX
  sources}\label{tab_photo_cdfn}
\scriptsize
\begin{tabular}{l cc cc cc cc cc c}
\hline ID & $U$ & $B$ & $R$ & $I$ & $z$ & $\rm 3.6\mu m$ & $\rm 4.5\mu
m$ & $\rm 5.8\mu m$& $\rm 8.0\mu m$ & $\rm 24\mu m$ & $\rm 70\mu m$\\ 
\hline 

irx-15   &   $-0.69\pm0.05$  &   $-0.52\pm0.04$   &  $0.11\pm0.01$   &
$0.59\pm0.01$ & $0.78\pm0.01$ & $1.91\pm0.01$ & $1.87\pm0.01$ & $1.84\pm0.01$ &
$1.80\pm0.02$ & $2.73\pm0.01$ & $3.24\pm 0.07$\\ 

irx-16&   $-0.70\pm0.04$   &    $-0.30\pm0.03$   &   $0.57\pm0.01$   &
$1.00\pm0.01$  &  $1.17\pm0.01$  &  $2.00\pm0.01$  &  $1.85\pm0.01$  &
$1.88\pm0.01$ & $1.89\pm0.02$ & $2.85\pm0.01$ & $4.09\pm 0.02$\\ 

irx-17&   $-0.83\pm0.05$   &   $-0.75\pm0.05$   &   $-0.04\pm0.01$   &
$0.39\pm0.01$  &  $0.54\pm0.01$  &  $1.49\pm0.01$  &  $1.37\pm0.01$  &
$1.39\pm0.04$ & $1.34\pm0.04$ & $2.47\pm0.01$ & $3.39\pm 0.05$\\ 

irx-18&   $-0.44\pm0.02$   &    $-0.33\pm0.04$   &   $0.22\pm0.01$   &
$0.57\pm0.01$  &  $0.71\pm0.01$  &  $1.73\pm0.01$  &  $1.86\pm0.01$  &
$2.07\pm0.01$ & $2.31\pm0.01$ & $2.99\pm0.01$ & $3.41\pm0.06$\\ 

irx-19&   $-0.94\pm0.08$   &    $-0.80\pm0.05$   &   $0.13\pm0.01$   &
$0.64\pm0.01$  &  $0.83\pm0.01$  &  $1.56\pm0.01$  &  $1.65\pm0.01$  &
$1.91\pm0.01$ & $2.12\pm0.01$ & $2.53\pm0.01$ & --\\

\hline
\end{tabular} 
\begin{list}{}{}
\item  None of  the sources  are  detected in  the GALEX  NUV and  FUV
  bands.  The columns  are: (1):  Source identification;  (2):  log of
  $U$-band flux density in units of $\rm \mu Jy$; (3): log of $B$-band
  flux density  in units of $\rm  \mu Jy$; (4): log  of DEEP2 $R$-band
  flux density  in units of $\rm  \mu Jy$; (5): log  of DEEP2 $I$-band
  flux density in  units of $\rm \mu Jy$; (6):  log of CFHTLS $z$-band
  flux density in units of $\rm  \mu Jy$; (7): log of IRAC $\rm 3.6\mu
  m$-band flux density in units of $\rm \mu Jy$; (8): log of IRAC $\rm
  4.5\mu m$-band  flux density in units  of $\rm \mu Jy$;  (9): log of
  IRAC  $\rm 5.8\mu m$-band  flux density  in units  of $\rm  \mu Jy$;
  (10): log of IRAC $\rm 8.9\mu  m$-band flux density in units of $\rm
  \mu Jy$; (11): log of MIPS  $\rm 24\mu m$-band flux density in units
  of $\rm \mu Jy$; (12): log of MIPS $\rm 70\mu m$-band flux density in units
  of $\rm \mu Jy$;
\end{list}
\end{table*}

\section{Sample Selection}

\begin{figure}
\begin{center}
 \rotatebox{0}{\includegraphics[height=0.9\columnwidth]{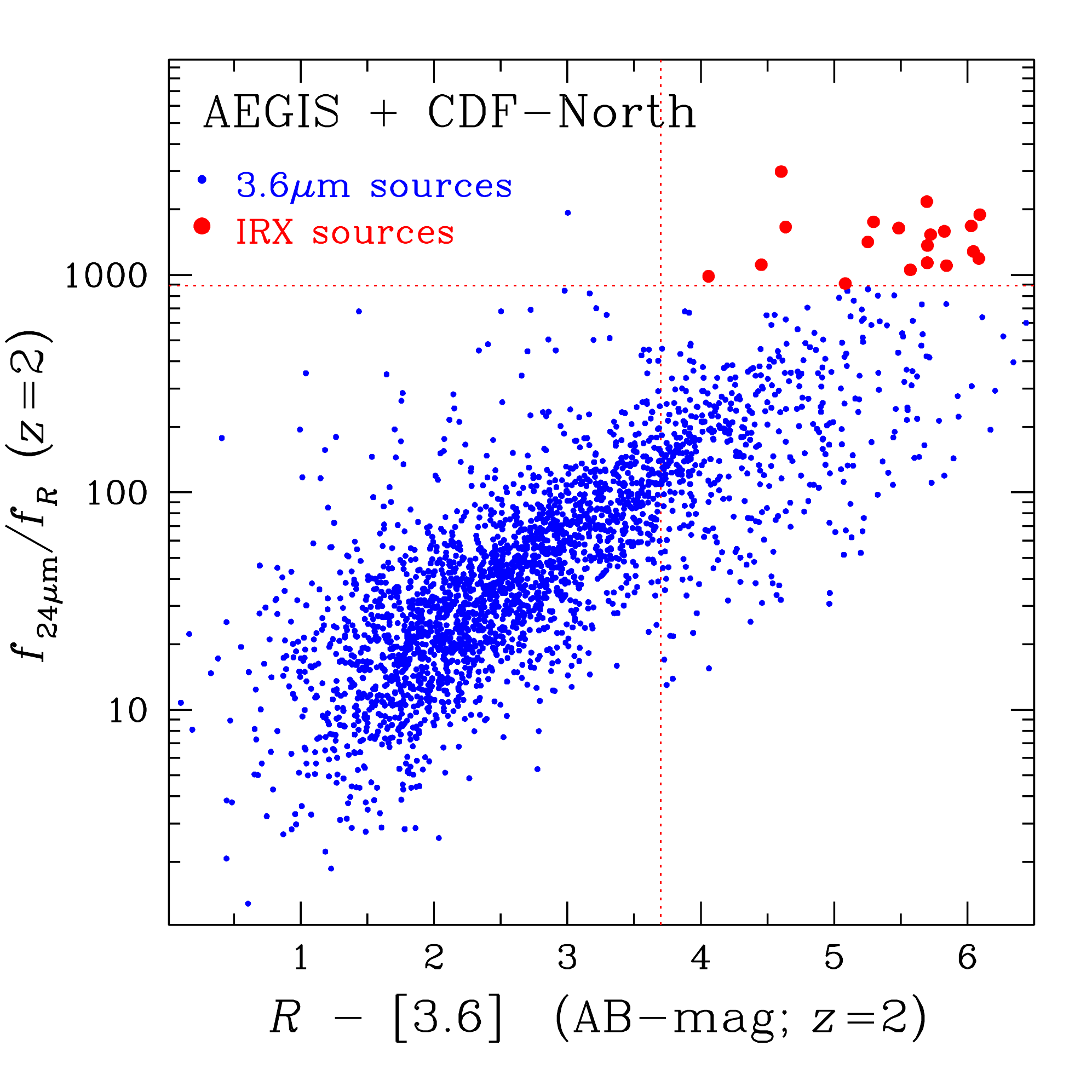}}
\end{center}
\caption{$f_{\rm 24\mu  m} / f_R$  against $R-\rm [3.6]$  estimated by
  redshifting to $z=2$ the rest-frame SEDs of $\rm 3.6\,\mu m$ sources
  in the  AEGIS and the  CDF-North with spectroscopic  redshifts (blue
  dots).  Large red  circles are the subset of  those sources which at
  $z=2$  have  $f_{\rm   24\mu  m}/f_{R}>900$  and  $R-[3.6]>3.7$\,mag
  (dotted horizontal and vertical lines respectively), i.e. similar to
  the  IRX  source  selection  criteria  (Fiore et  al.   2008,  2009;
  Georgantopoulos 2009).  }\label{fig_selection}
\end{figure}

Lower  redshift  analogues  of  the  $z\approx2$  IRX  population  are
identified as  follows. Firstly, IRAC  $\rm 3.6\, \mu m$  sources with
spectroscopic  redshifts  only  are  selected.  When  a  spectroscopic
quality   flag  is   available   we  select   sources  with   redshift
determinations secure  at the  $>90\%$ confidence level  (i.e. quality
flag $Q \ge  3$ for the DEEP2 and the TKRS  surveys). This reduces the
sample to 1784 sources in the GOODS-North and 6488 the AEGIS.  This is
to  avoid uncertainties in  the determination  of the  rest-frame SEDs
associated  with  insecure  spectroscopic redshift  determinations  or
photometric   redshift   errors.    The   rest-frame   SEDs   of   the
spectroscopically identified  IRAC $\rm 3.6\, \mu m$  sources are then
redshifted  to $z=2$,  the mean  redshift of  the IRX  population.  We
select only those sources which at $z=2$ have {\it observed} SEDs with
$f_{\rm 24\mu  m}/f_{R}>900$ and $R-[3.6]>3.7$\,mag  (AB system), i.e.
similar to the criteria  used by \cite{Fiore2008, Fiore2009} to define
their  IRX  sample.  \cite{Georgantopoulos2008}  have  shown that  the
colour  cut   $R-[3.6]>3.7$\,mag  adopted  here  is   similar  to  the
$R-K>4.5$\,mag  used by \cite{Fiore2008,  Fiore2009}.  The  method for
selecting lower  redshift analogues of the  $z\approx2$ IRX population
is  demonstrated by  the dotted  lines in  Figure \ref{fig_selection},
which  plots the flux  density ratio  $f_{\rm 24\mu  m}/f_{R}$ against
$R-[3.6]$ colour.  A total of 19 sources (14 in the AEGIS and 5 in the
GOODS-North)  are   selected.   The  sample  is   presented  in  Table
\ref{tab_sample}.   The observed  flux  densities of  the sources  are
shown  in   Table  \ref{tab_photo_aegis}  for  the   AEGIS  and  Table
\ref{tab_photo_cdfn} for the GOODS-North galaxies.

For the  estimation of the  $z=2$ SED we interpolate  linearly between
observed data  points. It is recognised  that the mid-IR  is a complex
wavelength  regime with PAHs  and/or the  silicate absorption/emission
feature at  $\rm 9.7\, \mu m$  having a strong impact  on the observed
broad-band  flux  densities.   The  linear interpolation  approach  is
ignoring these subtleties but  is preferred over more complex template
fitting  schemes  as  it   is  model  independent.   It  is  confirmed
nevertheless,  that the  sources presented  in  Table \ref{tab_sample}
satisfy the IRX criteria at $z=2$ if the SED fits described in section
\ref{sec_sed}  are  used  to  interpolate between  data  points.   The
Spitzer MIPS $\rm  24\,\mu m$ waveband and the  $R$-band correspond to
rest-frame wavelengths of  about $\rm 8\,\mu m$ and  $\rm 0.22\,\mu m$
at $z=2$ respectively.  In  order to avoid uncertain extrapolations of
the SED  we only consider  sources with flux density  detections, {\it
  not}  upper  limits, at  rest-frame  wavelengths  shorter than  $\rm
0.22\,\mu m$ and  longer than $\rm 8\,\mu m$.   In practice this means
that  for the  typical redshift  of the  sources in  the  sample, they
should be  detected at $\rm 24\,\mu  m$ or longer  wavelengths and the
$U$-band or shorter  wavelengths.  Three sources in the  AEGIS are not
detected at $\rm 24\,\mu m$ and therefore the observed $\rm 24\,\mu m$
flux density at $z=2$ is  estimated by interpolating the SED between 8
and $\rm 70\,\mu m$.

We caution that the sample of IRX sources selected in this paper is by
no  means   complete.   For   example,  the  requirement   for  secure
spectroscopic redshifts  translates to  an optical magnitude  limit of
$R\approx24$\,mag.   It  can be  shown  however,  that  this does  not
introduce a  bias to the sample,  i.e.  in favour  of optically bright
sources.  We  calculate that the  median {\it observed}  $R$-band flux
density of  our sources if  they were at  $z=2$ would be  $\rm \approx
0.04 \, \mu Jy$.  This is estimated after scaling with the appropriate
k-corrections  (i.e.   $1+z$  factor)  and  luminosity  distances  the
$B$-band flux  densities of our  sources, which roughly  correspond to
the $R$-band at  $z=2$.  The number above should  be compared with the
median  $R$-band flux  density of  $\rm \approx  0.07 \,  \mu  Jy$ for
sources in the AEGIS and  the GOODS-North fields with observed $f_{\rm
  24\mu m}/f_{R}>900$ and $R-[3.6]>3.7$\,mag (AB system).  If anything
our sample is sensitive to somewhat optically fainter sources compared
to the  $z\approx2$ IRX  population in the  AEGIS and  the GOODS-North
fields. Additionally, the median $\rm  5.8\mu m$ luminosity of our IRX
sample,  is $\nu L_{\nu}(\rm  5.8\mu m)\approx1.6\times10^{44}  \, erg
\,s^{-1}$,  about 2  times  fainter  than the  median  $\rm 5.8\mu  m$
luminosity of the IRX sources in the CDF-South ($\rm S_{24}>40\mu Jy$;
Fiore et al.  2008) and 1\,dex  fainter than those in the COSMOS ($\rm
S_{24}>500\mu Jy$;  Fiore et al.   2009).  We therefore  conclude that
our  $z\approx1$ sample,  although  not complete,  is  similar to  the
$z\approx2$ IRX  sources found  in the CDF-South,  for which  Fiore et
al. (2008) estimate a Compton Thick  rate of $80\pm15$ per cent and an
intrinsic  luminosity of  $L_X(\rm  2-10\,keV) >  10^{43}  \rm erg  \,
s^{-1}$.

\begin{figure*}
\begin{center}
\rotatebox{0}{\includegraphics[height=1.7\columnwidth]{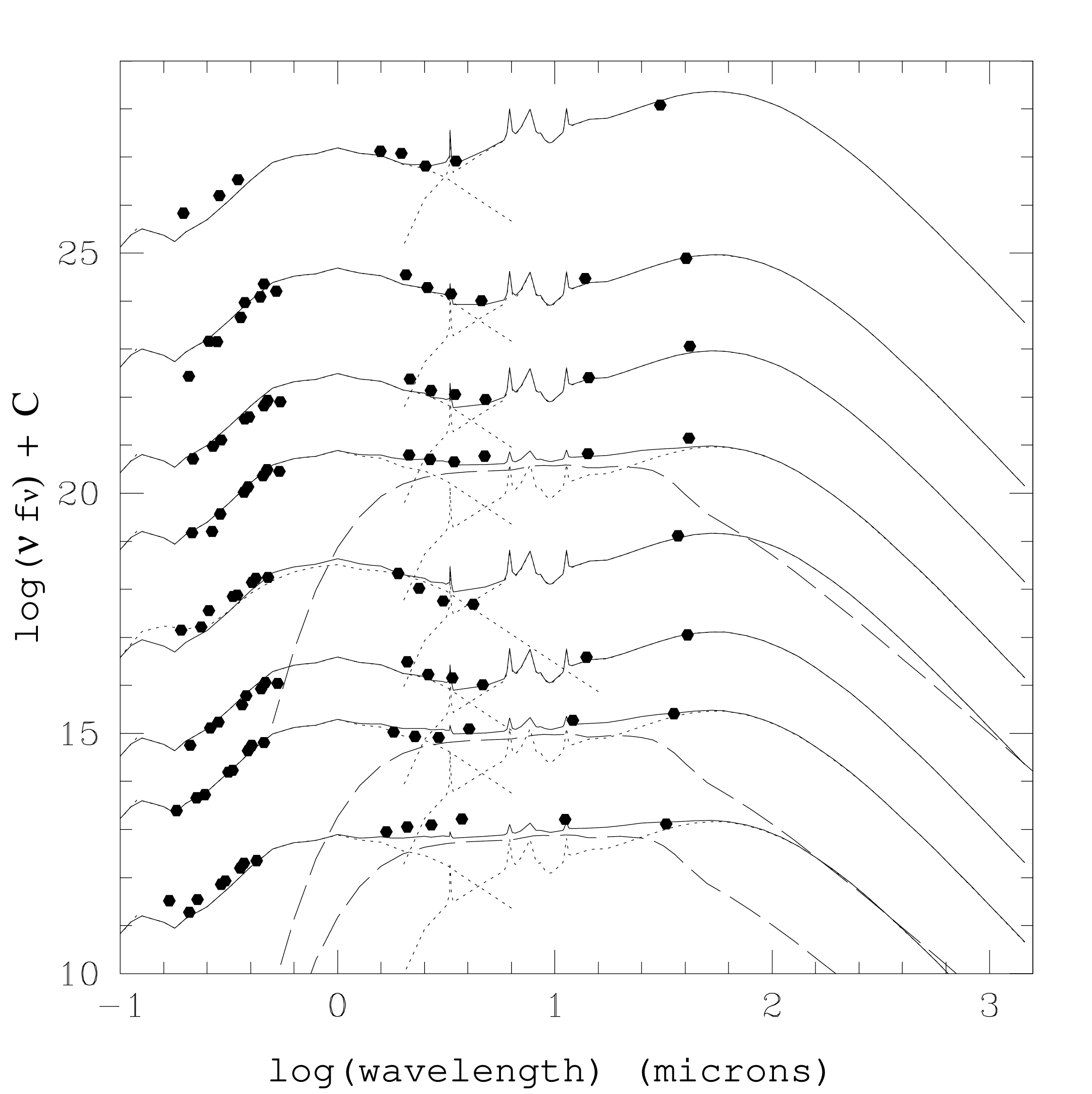}}
\end{center}
\caption{Examples of  template fits  to the SEDs  of the  IRX selected
  galaxies  in  the AEGIS.   From  bottom  to  top are:  irx-8  (X-ray
  source),  irx-9  (X-ray   source),  irx-10,  irx-11,  irx-12  (X-ray
  source),  irx-13,  irx-14 and  irx-7.   The  dots  are the  observed
  UV--to--far-IR photometry and the  continuous lines are the best-fit
  models.  Different  sources are offset by an  arbitrary constant for
  clarity.   The  dashed  curves  correspond  to the  hot  dust  torus
  component,  which is  needed  for all  X-ray  detected sources  with
  $L_X(\rm 2-10\,keV) \ga 10^{43} \,  erg \, s^{-1}$. The dotted lines
  at    infrared   wavelengths    are   the    starburst   components.
}\label{fig_sed_aegis}
\end{figure*}

\begin{figure*}
\begin{center}
\rotatebox{0}{\includegraphics[height=1.7\columnwidth]{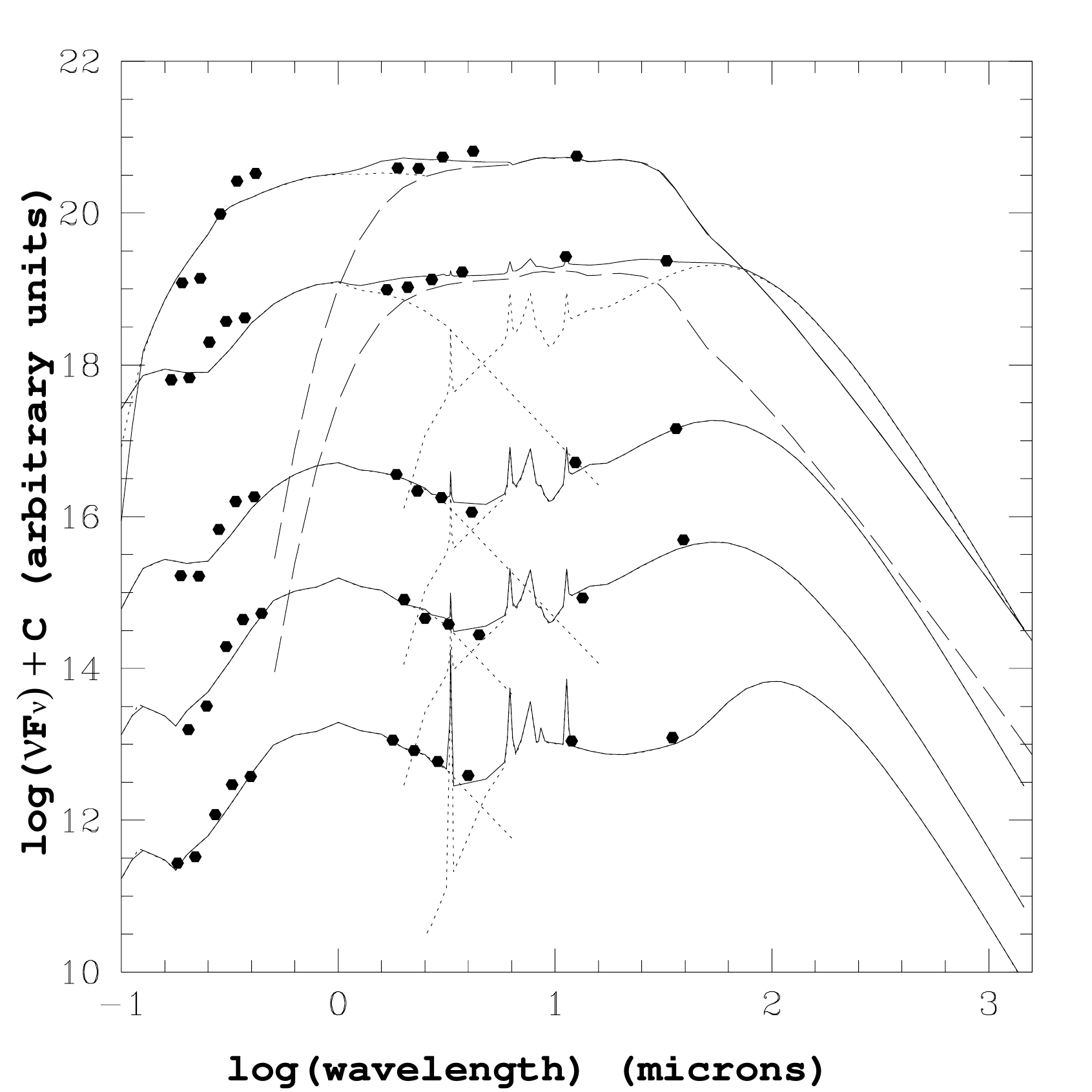}}
\end{center}
\caption{Same  as  in  figure  \ref{fig_sed_aegis}  for  IRX  selected
  galaxies in the CDF-North.  From  bottom to top are: irx-15, irx-16,
  irx-17,  irx-18, irx-19.  The  curves and  dots are  the same  as in
  Figure \ref{fig_sed_aegis}.  }\label{fig_sed_hdfn}
\end{figure*}

\begin{figure}
\begin{center}
\rotatebox{0}{\includegraphics[height=0.9\columnwidth]{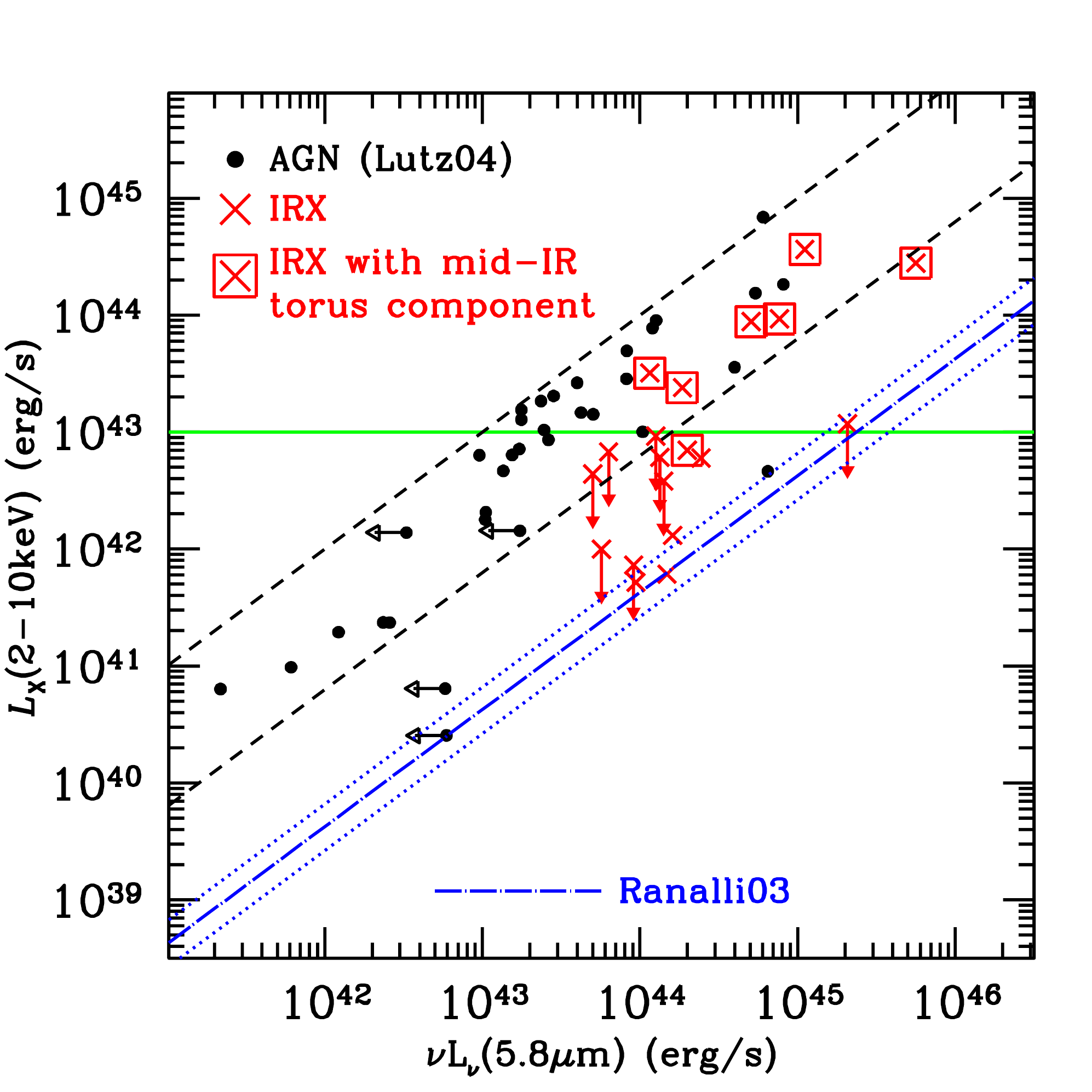}}
\end{center}
\caption{X-ray $\rm 2-10 \,keV$ luminosity plotted against $\rm 5.8\mu
  m$ luminosity. Filled circles (black)  are local AGN from the sample
  of Lutz et al.  (2004).  The dashed lines show the dispersion around
  the mean  $L_X({\rm 2-10\,keV})-\nu  L_{\nu} (\rm 5\mu  m)$ relation
  for  the Lutz et  al.  (2004)  AGN.  The  (blue) long-dashed--dotted
  line is  the $L_X  - \nu L_{\rm  5\mu m}$ relation  for star-forming
  galaxies adapted from (Ranalli et al. 2003; see Fiore et al.  2009).
  The dotted lines correspond to the 1\,sigma rms envelope around this
  relation.   Our  sample of  IRX  sources  is  shown with  the  (red)
  crosses.  For IRX sources that are not detected at X-ray wavelengths
  the upper limit in $L_X(2-10\,keV)$ is plotted. A square on top of a
  cross signifies IRX  sources which require a torus  component to fit
  their infrared  SEDs.  Above an  intrinsic X-ray luminosity  of $\rm
  10^{43}\,erg\,s^{-1}$ (shown with the horizontal green line) all IRX
  sources with  an X-ray counterpart  in the 2-10\,keV band  require a
  torus component.}\label{fig_ll}
\end{figure}

\section{SED fitting}\label{sec_sed}

The observed  optical to mid-IR Spectral Energy  Distribution (SED) of
the sample  sources are modeled following the  methods fully described
in  \cite{rowan2005,rowan2008}.   In  brief  the  $FUV$-band  to  $\rm
4.5\,\mu m$  photometric data are fit  using a library  of 8 templates
described by  \cite{Babbedge2004}, 6 galaxies  (E, Sab, Sbc,  Scd, Sdm
and sb) and  2 AGN.  At longer  wavelengths ($\rm 5.8 - 24  \, \mu m$)
any dust  may significantly contribute  or even dominate  the observed
emission.   Before fitting  models  to these  wavelengths the  stellar
contribution is subtracted from  the photometric data by extrapolating
the best-fit  galaxy template from  the previous step.   The residuals
are   then   fit   with   a   mixture  of   four   templates:   cirrus
\citep{Efstathiou2003},     AGN     dust    tori     \citep{rowan1995,
  Efstathiou1995},       M\,82      and       Arp\,220      starbursts
\citep{Efstathiou2000}. The  modeling above provides  both information
on the  dominant emission  mechanism in the  optical and  the infrared
(AGN  vs  star-formation)  and  an  estimate  of  the  total  infrared
luminosity, $L_{TOT}$, in the  wavelength range $\rm 3-1000\mu m$.  As
discussed  by \cite{rowan2005}  $L_{TOT}$ is  expected to  be accurate
within a factor of two.  In this exercise the redshift is fixed to the
spectroscopically determined value.

The  results  are presented  in  Table  \ref{tab_sample}. Examples  of
template  fits to  SED of  AEGIS and  CDF-North sources  are  shown in
Figures \ref{fig_sed_aegis} and \ref{fig_sed_hdfn}. Our analysis shows
that  dust associated with  star-formation (M\,82  starburst template)
either  dominates or  substantially  contributes to  the  flux at  the
mid/far-infrared wavelengths of most  sources.  In this respect, it is
not surprising  that 12/14  galaxies in AEGIS  and 4/5 sources  in the
CDF-N  are   detected  at  this  longer   wavelength.   The  estimated
star-formation rates  are in the range  $\rm 10-600\,M_{\odot}/yr$. The
best-fit  starburst  template is  that  of  M\,82,  in agreement  with
results from  \cite{Bussmann2009_sed} who showed that  the far-IR SEDs
of  IRX sources  are  better described  by  warm-dust templates  (e.g.
similar  to M\,82  or Mrk\,231)  and are  inconsistent with  cold dust
templates (e.g. Arp\,220).

An additional QSO  torus component is required to  fit the mid-IR part
of the SED of some sources in Table \ref{tab_sample}. As will be shown
in the  next section this component  is needed for  only those sources
which are detected at X-ray wavelengths and which have intrinsic X-ray
luminosities, $L_X(\rm 2-10\, keV) \ga 10^{43} \, erg \, s^{-1}$.  The
SED  fitting approach  can  therefore  be used  to  identify AGN  with
intrinsic luminosity $\rm  \ga 10^{43} \, erg \,  s^{-1}$.  Below this
limit, star-formation in the host  galaxy may swamp any AGN signatures
in the mid-IR.

\begin{table*}
\caption{X-ray properties of the sample}\label{tab_xspec}
\begin{tabular}{lccc ccc ccc}
\hline ID  & XID & counts  & bkg & $\rm  N_H$ & $R$ &  $\Gamma$ & $\rm
c-stat$ & dof & $L_X(\rm 2-10\,keV)$\\ & & & & ($\rm \times 10^{22} \,
cm^{-2}$)& & &  & & ($\rm \times 10^{43} \, erg  \, s^{-1}$) \\ \hline
irx-1 & -- & 16 & 7.3 & -- & -- & -- & -- & --& $<0.7$\\
     
irx-2   &  egs\_0263   &  127   &  0.6   &   $44.6_{-12.1}^{+26.0}$  &
$0.6_{-0.6}^{+1.2}$ & 1.9(f) &42.1 & 46 & 36.4 \\

irx-3 & -- & 0 & 0.2 & -- & -- & -- & -- & --& $<0.07$\\
        
irx-4 i & -- & 4 & 0.5 & -- & -- & -- & -- & --& $0.1$\\


irx-5   &   egs\_0311   &   112   &  5.9   &   $7.9_{-2.5}^{+2.9}$   &
$2.3_{-2.9}^{+8.0}$ & 1.9(f) &40.5 & 46 & 3.2 \\

irx-6 &-- & 14 &10.4 & -- & -- & -- & -- & --& $<0.6$\\

irx-7 & -- & 2 & 0.2 & -- & -- & -- & -- & --& $<0.9$\\

irx-8    &    egs\_0416&    820    &   1.0    &    $2.1_{-0.8}^{+0.6}$
&$0.03_{-0.03}^{+7.8}$& $2.1^{+0.7}_{-0.2}$ &48.2 & 45 & 27.9 \\

irx-9   &   egs\_0424   &   254   &  1.2   &   $4.6_{-1.1}^{+1.1}$   &
$0.5_{-0.5}^{+2.7}$ & 1.9(f) &44.9 & 46 & 8.8 \\

irx-10 &-- & 0 & 0.1 & -- & -- & -- & -- & --& $<0.1$\\

irx-11 &-- & 6 &4.1 & -- & -- & -- & -- & --& $<0.4$\\

irx-12   &  egs\_0917   &  163   &  47.0   &   $10.3_{-3.5}^{+3.5}$  &
$2.6_{-2.6}^{+10.7}$ & 1.9(f) & 43.4 & 46 & 2.4 \\


irx-13 &-- & 19 & 11.8 & -- & -- & -- & -- & --& $<0.4$\\

irx-14 &-- & 2 & 0.2 & -- & -- & -- & -- & --& $<1.2$\\

               & & & & & \\

irx-15   &   hdfn\_098   &   174   &28.0   &   $9.5_{-4.6}^{+5.7}$   &
$6.4_{-6.4}^{+22.8}$ & 1.9(f) & 41.4 & 39 & 0.6 \\

irx-16 & hdfn\_328 & 129 & 96.7&$1.3_{-1.3}^{+4.1}$ & -1(f) & 1.9(f) &
50.3 & 47 & 0.05 \\

irx-17 & hdfn\_466 & 27 & 20.5 &$<2.8$  & -1(f) & 1.9(f) & 43.1 & 49 &
0.046 \\

irx-18   &   hdfn\_023   &   2240  &   74.2&   $2.3_{-0.6}^{+0.8}$   &
$5.6_{-2.5}^{+22.5}$ & $2.0^{+0.7}_{-0.3}$ &51.4 & 45 & 9.3 \\

irx-19   &   --   &   260   &  210.1   &   $55.4_{-38.2}^{+117.5}$   &
$0.2_{-0.2}^{+18.4}$ & 1.9(f) &103.5 & 117 & 0.7 \\

\hline

\end{tabular} 
\begin{list}{}{}
\item  The  columns  are:   (1):  Source  identification;  (2):  X-ray
  identification  number  as in  Laird  et  al  (2009) for  AEGIS  and
  Georgakakis et  al.  (2008) for  the CDF-North.  Sources  without an
  X-ray   counterpart  or   low   significance  (Poisson   probability
  $>4\times10^{-6}$) X-ray sources do not have an X-ray identification
  number  (3): total  counts  in the  energy  range 0.5-10\,keV.   For
  sources with insufficient counts  to perform X-ray spectral analysis
  the total  extracted counts in  the 2-8\,keV band are  listed.  (4):
  background  counts  in  the  range 0.5-10\,keV.   For  sources  with
  insufficient   counts  to  perform   X-ray  spectral   analysis  the
  background counts  in the 2-8\,keV  band are listed.   (5): best-fit
  column density of the cold absorber (WABS component of XSPEC model).
  The uncertainties  correspond to the  90 per cent  confidence level;
  (6): reflection fraction  of the PEXMON model of  XSPEC.  The errors
  correspond  to the  90  per cent  confidence  level; (7):  power-law
  photon  index, $\Gamma$,  of the  intrinsic AGN  power-law spectrum.
  The ``(f)'' indicates that this parameter was kept fixed in the fit.
  (10): C-stat of the best-fit solution; (9): degrees of freedom; (11)
  obscuration corrected luminosity in the 2-10\,keV range and in units
  of $\rm 10^{43}  \, erg \, s^{-1}$.  In the case  of upper limits or
  sources with insufficient counts to perform X-ray spectral analysis,
  no correction has been  applied for intrinsic obscuration.  In those
  sources fluxes and luminosities  in the 2-10\,keV band are estimated
  by   adopting  a   power-law  spectral   index   with  $\Gamma=1.4$,
  i.e. assume that  the sources have X-ray spectra  similar to that of
  the X-ray background. The  statistical uncertainty in the luminosity
  is less  than 10\% for  X-ray detected sources  (Poisson probability
  $<4\times10^{-6}$)  and about  a factor  of 2  for  low significance
  ones.
\end{list}
\end{table*}

\section{X-ray properties}\label{sec_results}

\subsection{X-ray source counterparts}

The  cross-correlation  between the  X-ray  and  IRAC  $\rm 3.6\mu  m$
catalogues using the  Likelihood Ratio method shows that  5/16 and 4/5
of the  sample sources in  AEGIS and the CDF-North  respectively, have
X-ray counterparts  with Poisson significance  $<4\times10^{-6}$ in at
least  one  of   the  4  energy  bands  used   for  detection.   Lower
significance X-ray counterparts are also searched for in the hard band
(2-7\,keV), which provides a reasonable compromise between sensitivity
and obscuration effects,  thereby minimising biases when extrapolating
fluxes  to the rest-frame  2-10\,keV energy  range.  The  X-ray counts
within the 70  per cent Encircled Energy Fraction  (EEF) radius at the
position of each source in Table 1 are summed up and a local value for
the background  is estimated using  an annulus centered on  the source
with  inner  radius  1.5 times  the  90  per  cent  EEF and  width  of
50\,arcsec (100  pixels).  For the estimation of  the background X-ray
sources  with Poisson  significance $<4\times10^{-6}$  are  removed by
excluding pixels  within the  95 per cent  EEF radius of  each source.
The probability that  the observed counts are a  random fluctuation of
the background is estimated  using Poisson statistics.  We consider as
detections sources with  Poisson significance $<3\times10^{-3}$.  This
exercise yields 2 lower significance  X-ray detections, 1 in the AEGIS
and 1 in the CDF-N.  The fluxes of these sources are estimated using a
Bayesian     methodology    similar     to    that     described    in
\cite{Laird2009}. For  a source with  observed number of  total counts
$N$ (source  and background) within the  70 per cent EEF  radius and a
local background value $B$ the probability of flux $f_X$ is

\begin{equation}\label{eq_source} P(f_X, N) = \frac{T^N \, e^{-T}}{N!}
\, \pi(f_X),
\end{equation}

\noindent where  $N=S+B$ and $S =  f_X \times t_{exp}  \times C \times
\eta$.   In the  last equation  $t_{exp}$ is  the exposure  time  at a
particular position after accounting  for instrumental effects, $C$ is
the  conversion factor  from  flux to  count  rate and  $\eta$ is  the
encircled energy fraction, i.e. 0.7  in our case.  The term $\pi(f_X)$
in  equation  \ref{eq_source}  accounts  for the  Eddington  bias  and
assumes  that the  differential X-ray  source counts  follow  a broken
power-law for the  $\log N - \log S$ with  faint and bright-end slopes
of --1.5 and --2.5 respectively and  a break flux of $\rm 10^{-14} erg
\, s^{-1}  \, cm^{-2}$ \citep{Georgakakis2008_sense}.   For the source
flux we adopt the mode of the distribution $P(f_X, N)$.


The same methodology is also  adopted to estimate upper limits for the
fluxes of the sources that  are not detected at X-ray wavelengths. The
confidence interval CL is given by the integral

\begin{equation}\label{eq_limit} \int_{f_{X,L}}^{f_{X,U}} P(f_X, N) \,
df_x=CL.
\end{equation}

For a faint-end  slope of --1.5 the integral  diverges at the faintest
fluxes.  To  avoid this  problem we  adopt as the  lower limit  of the
integration  $f_{X,L}= \rm  10^{-18}  \, erg  \,  s^{-1} \,  cm^{-2}$.
Equation  \ref{eq_limit} is  then solved  numerically to  estimate the
flux upper  limit $f_{X,U}$ for  the confidence interval of  99.97 per
cent. The results are presented in Table 1.

\subsection{X-ray spectral analysis}

The X-ray spectra of the  sample sources were extracted using the ACIS
extract \citep[AE; v. 2008-03-04; ][]{acis_extract} IDL package, which
is  designed  to  deal  with  large  number  of  sources  on  multiple
observations.   The AE  package User's  Guide is  available  online at
{http://www.astro.psu.edu/xray/docs/TARA/ae\_users\_guide.html}.    The
procedures  used  in  AE  are  described  in  \cite{Townsley2003}  and
\cite{Getman2005}. The X-ray counts of each source were extracted from
individual  pointings using the  95 per  cent encircled  energy radius
(1.5\,keV)  at  the position  of  the  source.   The composite  source
spectrum  was  then  constructed  by  summing  the  counts  from  each
observation.  For the background  estimation, point sources were first
masked out  using a circular region  with size 1.1 larger  than the 99
per cent EEF at 1.5\,keV.   The background spectra were then extracted
from  the  masked event  files  of  individual  observations using  an
annulus centered  on the  source with variable  size that  enclosed at
least 150 background counts.  The spectra from separate pointings were
scaled by the ratio of the  exposure time in the source and background
regions and then summed up  into the composite background spectrum for
a  particular  source.   The  Response  Matrix Files  (RMF)  and  Area
Response  Files  (ARF) for  individual  observations were  constructed
using  the   {\sc  ciao}  tools   {\sc  mkacisrmf}  and   {\sc  mkarf}
respectively. These were then combined into composite RMFs and ARFs by
the  {\sc ftool}  tasks {\sc  addrmf} and  {\sc  addarf} respectively,
using the exposure time of  each observation as weight. Source spectra
were grouped to have at least one count per bin.

For the spectral  fits we use the XSPEC v12. We  adopt a realistic AGN
spectral model to fit the  data, motivated by current observations and
ideas about  the structure  of the regions  close to SBHs.   The model
consists  of power-law  absorbed  by cold  material  and a  reflection
component as described below.

The main  spectral component of AGN  is the power law  with an average
photon  index $\Gamma=1.9$  \citep{Nandra1994}  and a  cutoff at  high
energies  \citep[$\approx  100$\,keV;][]{Gondek1996}.  The  dispersion
around   the    mean   $\Gamma$    is   estimated   to    be   0.2-0.3
\citep{Gondek1996}.   The reflection  component represents  direct AGN
emission  that  intersects  optically  thick  material,  such  as  the
accretion disk  or the  torus, and is  Compton backscattered  into the
line of sight.   We parametrise this component using  the {\sc pexmon}
model  of  XSPEC  \citep{Nandra2007_Fe},  which includes  an  analytic
formulation  for the  strength of  the narrow  iron line  at 6.4\,keV,
which is  expected to  be prominent in  the reflection  spectrum.  The
direct emission of AGN is also modified by cold material. We adopt the
photoelectric  absorption cross  sections  of \cite{Morrison1983}  for
solar metallicity  to represent any  cold absorbing medium, such  as a
torus.   Ideally,  our  modeling  should also  include  absorption  by
ionised material. However, given the relatively small number of counts
in the typical  X-ray spectra of deep field sources,  we choose not to
include this component in the  analysis.  In any case, warm absorption
is  predominantly affecting  the  AGN spectra  at  soft energies  $\la
1$\,keV  \citep[e.g.][]{George1998}.   At   least  part  of  the  warm
absorber spectral  features are  therefore, expected to  be redshifted
out  of the  {\it  Chandra}  observable energy  range  for sources  at
moderate and  high redshifts.  We also perform  the spectral modelling
for source  photons $>2$\,keV (observer's  frame) to ensure  that soft
emission associated with  a warm absorber does not  modify the results
and conclusions.  In summary the  XSPEC model we are using consists of
a power-law absorbed  by cold gas at the rest-frame  of the source and
Compton reflection.  These components are also absorbed by cold gas in
our Galaxy  using the appropriate  Galactic HI column density  for the
AEGIS ($\rm N_H=1.4\times10^{20} \,  cm^{-2}$) and the CDF-North ($\rm
N_H=1.3\times10^{20} \, cm^{-2}$).  In XSPEC terminology, our model is
$\mathtt{wabs}    \times   (\mathtt{zwabs}   \times    \mathtt{po}   +
\mathtt{pexmon})$, where  $\mathtt{wabs}$ and $\mathtt{zwabs}$  is the
cold absorption from  our Galaxy and at the rest  frame of the sources
respectively, and $\mathtt{po}$ is the power-law.

The parameters of  the model are the photon index  and the high energy
cutoff (pexmon only) of the direct AGN emission, the column density of
the cold absorber, $N_H$,  the element abundances, the inclination $i$
of the slab that gives the reflection and the reflection fraction $R$,
defined as  the strength of  the reflection relative to  that expected
from a slab subtending a solid angle of $2\pi$. Clearly, these are too
many free  parameters given the number  of counts in  our spectra (see
Table \ref{tab_xspec}).   We therefore fix abundances  to solar values
and the  inclination angle to  60\,deg.  Given that the  spectra cover
the energy range 2-10\,keV  (rest frame 4-20\,keV at $z\approx1$), the
energy cutoff of  the direct AGN emission will  have negligible impact
on the results.  We therefore  choose to fix this parameter to 1000keV
(pexmon only).   These assumptions leave 3  free parameters, $\Gamma$,
$N_H$ and $R$.  For X-ray sources in the sample with $<200$ net counts
in the 0.5-10\,keV range we also fix the photon index of the power-law
to $\Gamma=1.9$,  while for sources  with $<50$ net counts  we further
fix the reflection fraction to $R=1$.

In the case of Compton Thick  AGN we expect either to see directly the
turnover  in the  spectrum at  rest-frame energies  $\rm  \ga 15\,keV$
because of  the cold absorption  (i.e. estimate a column  density $\rm
N_H  > 2  \times 10^{24}\,cm^{-2}$  from the  spectral fit)  or  get a
reflection dominated spectrum with $R>>10$ \citep{Nandra2007}.

\begin{figure*}
\begin{center}
\rotatebox{270}{\includegraphics[height=0.8\columnwidth]{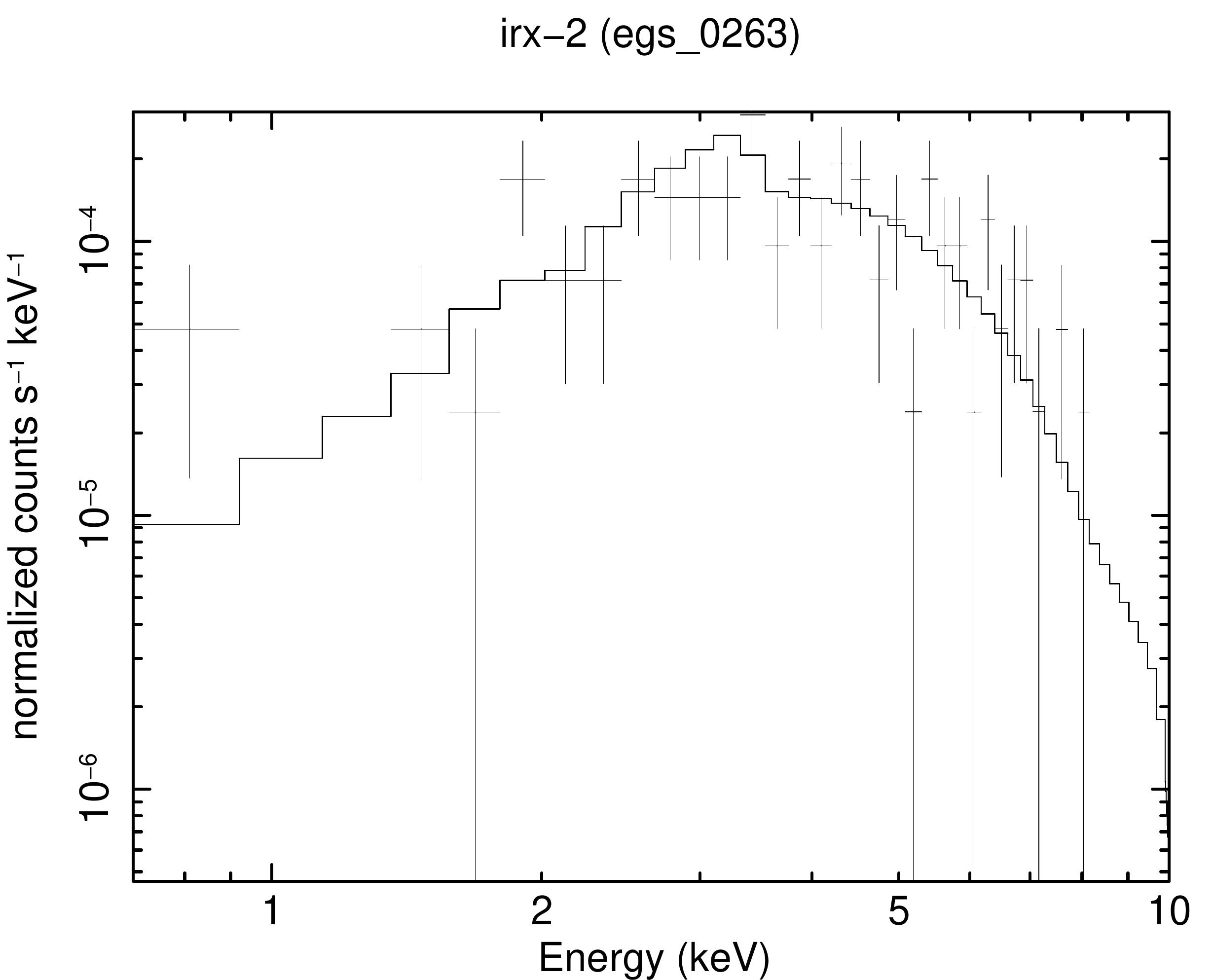}}
\rotatebox{270}{\includegraphics[height=0.8\columnwidth]{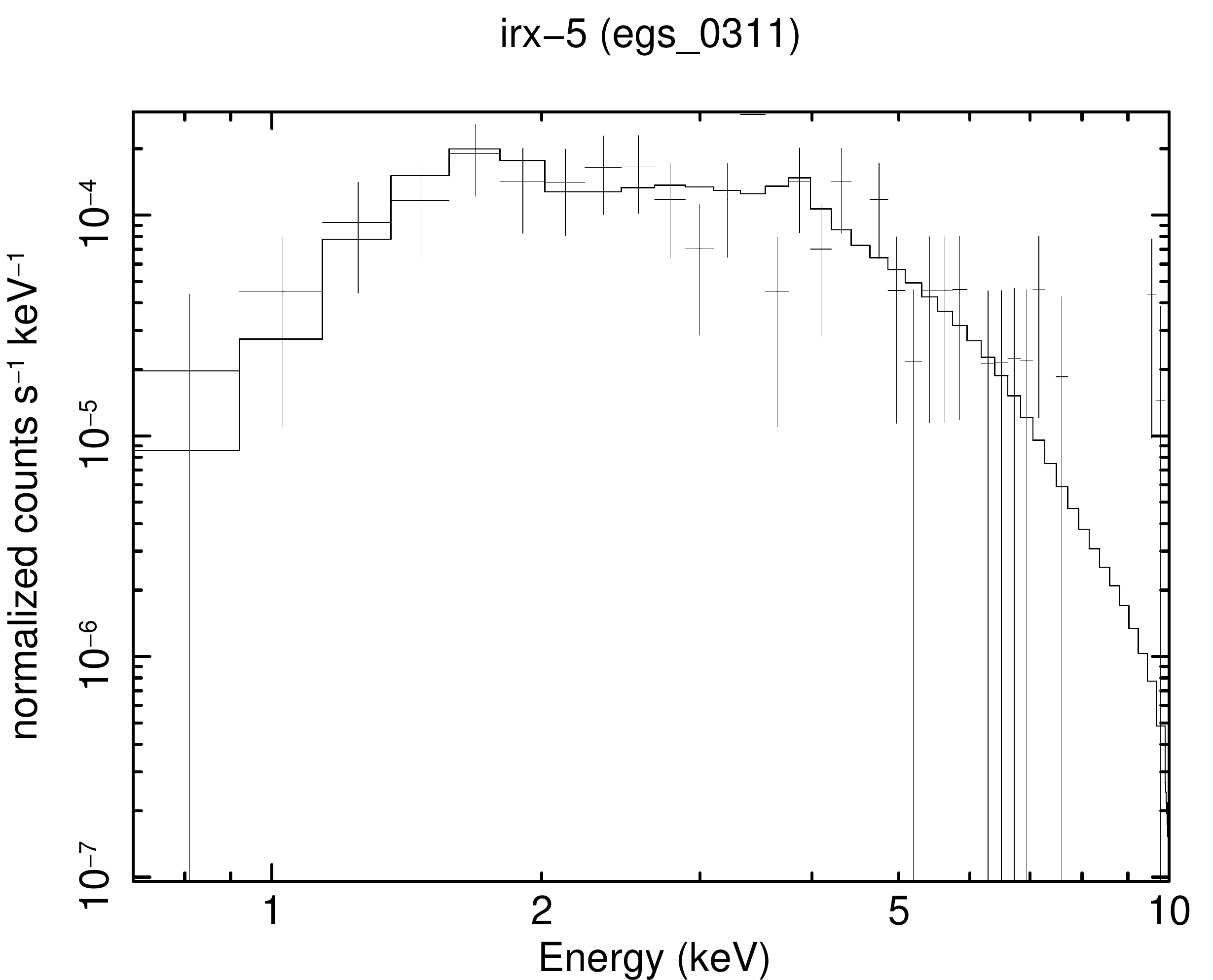}}
\rotatebox{270}{\includegraphics[height=0.8\columnwidth]{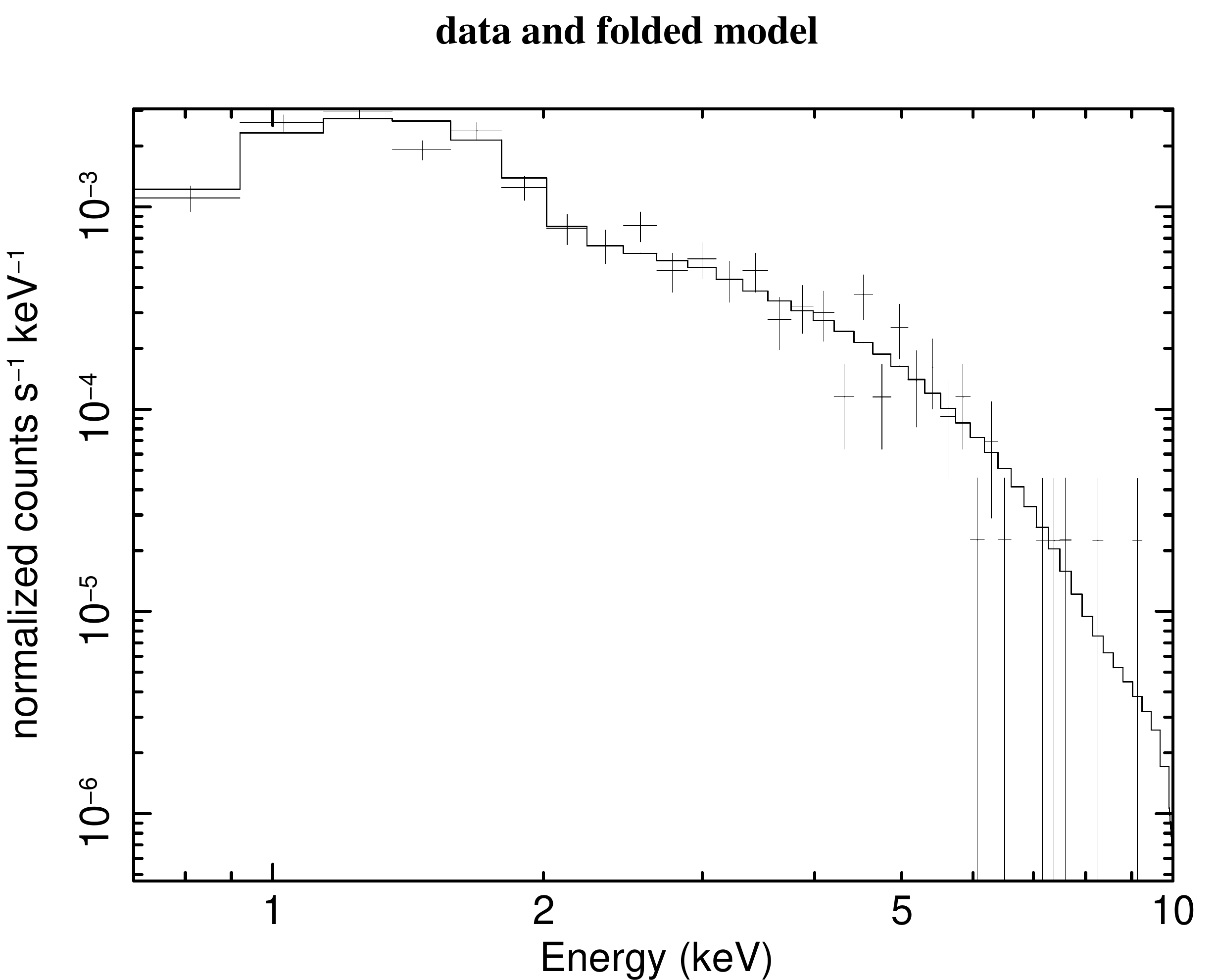}}
\rotatebox{270}{\includegraphics[height=0.8\columnwidth]{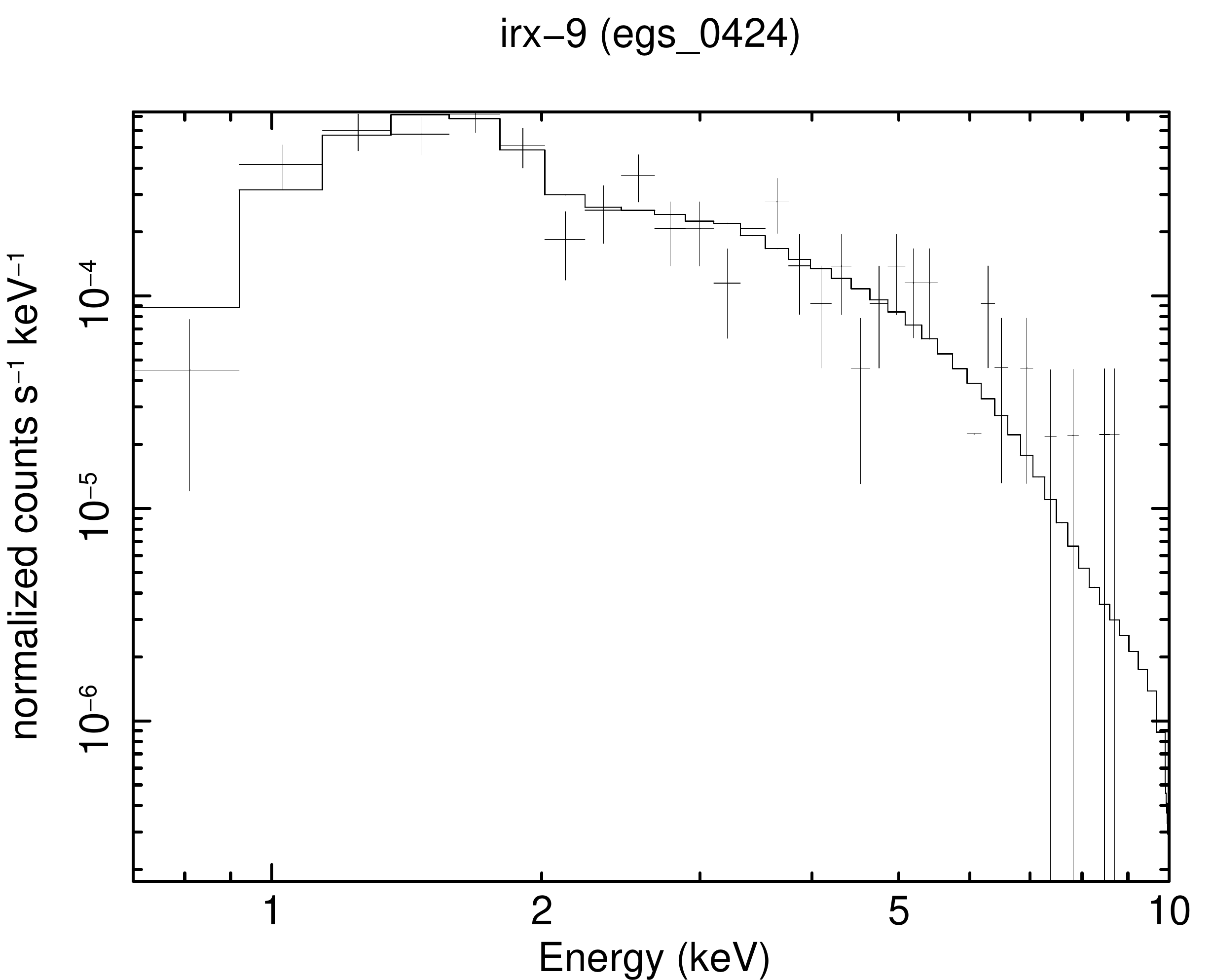}}
\rotatebox{270}{\includegraphics[height=0.8\columnwidth]{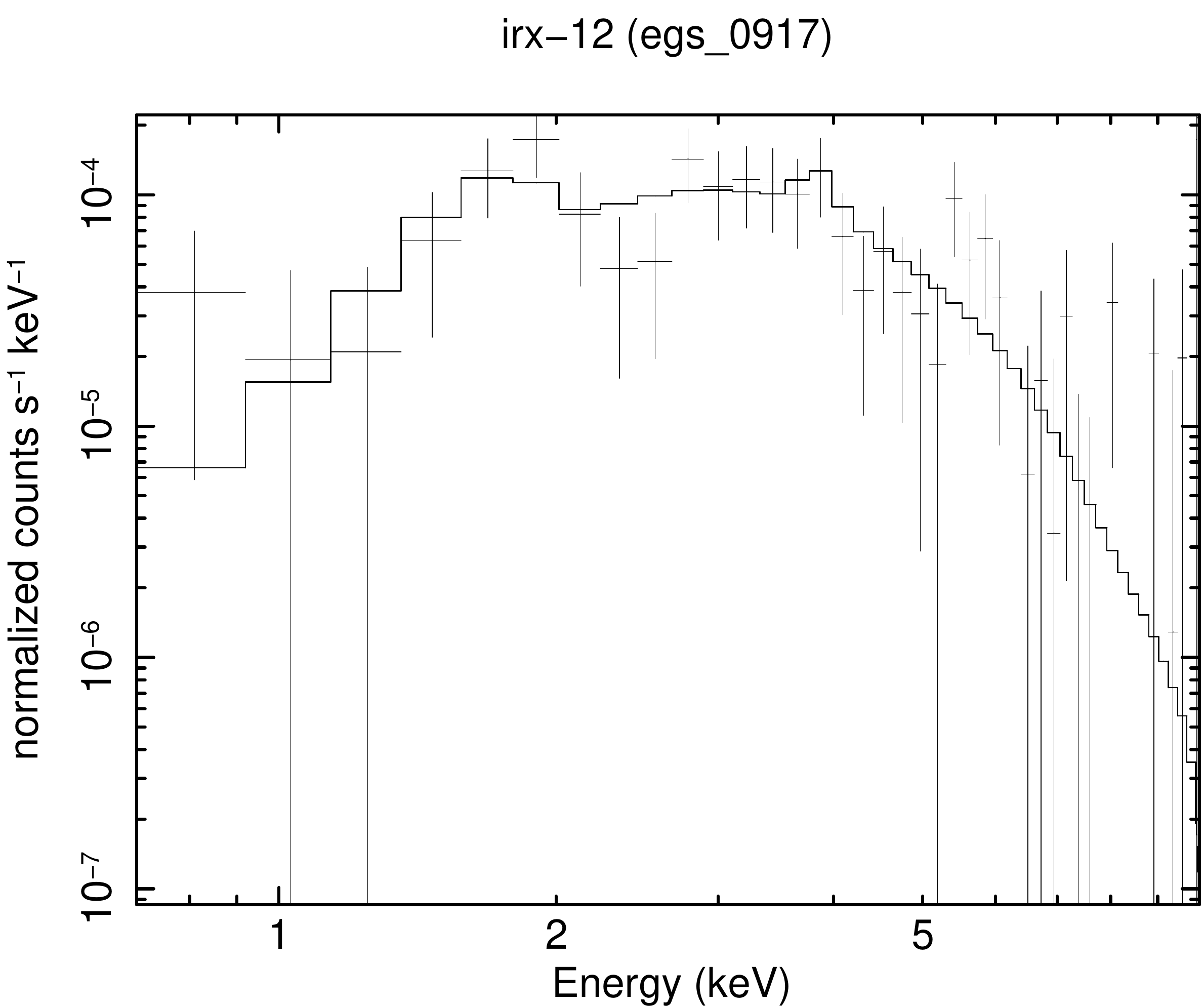}}
\end{center}
\caption{X-ray spectra  and spectral fits  for the IRX sources  in the
  AEGIS.  }\label{fig_xspec1}
\end{figure*}

\begin{figure*}
\begin{center}
\rotatebox{270}{\includegraphics[height=0.8\columnwidth]{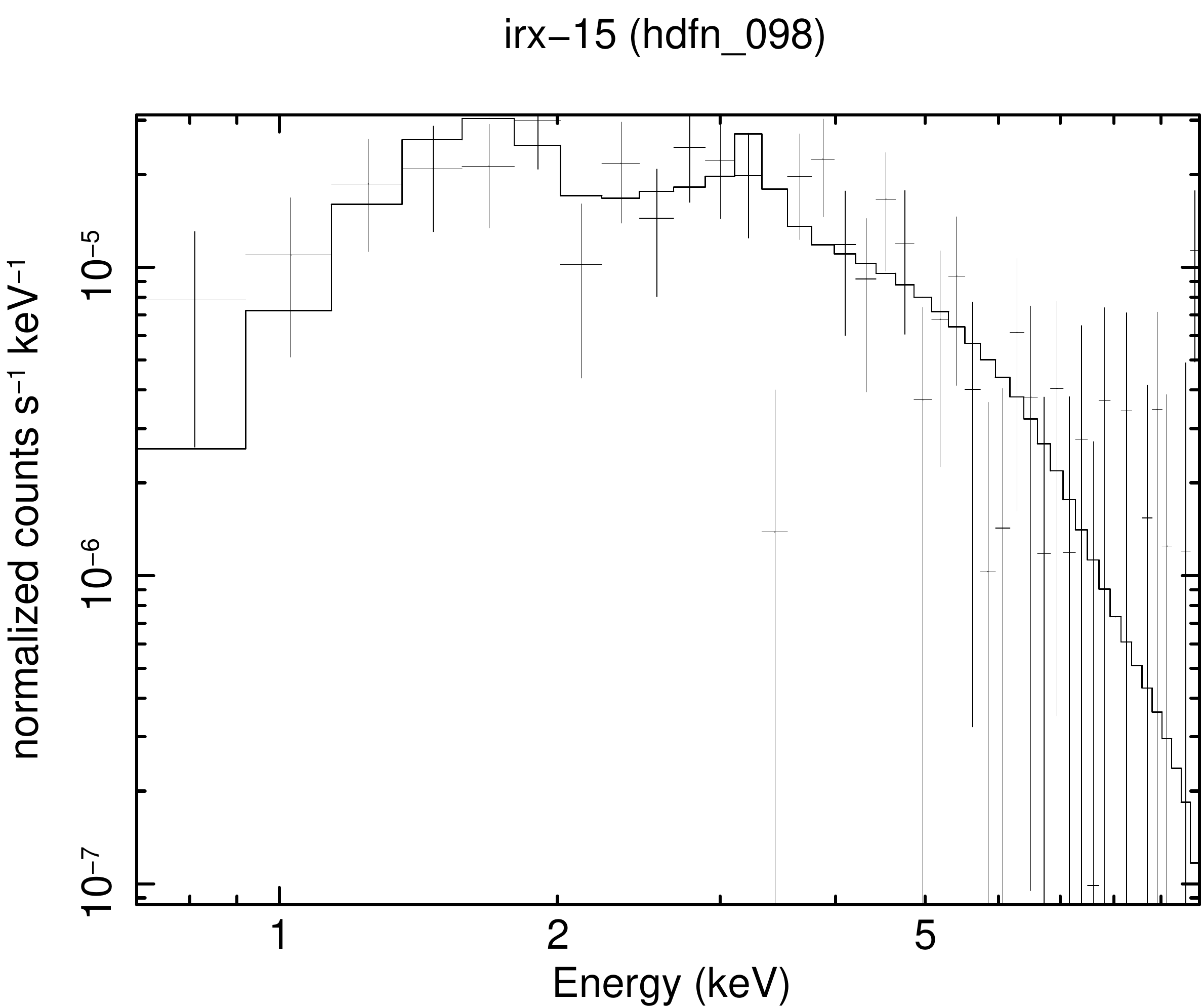}}
\rotatebox{270}{\includegraphics[height=0.8\columnwidth]{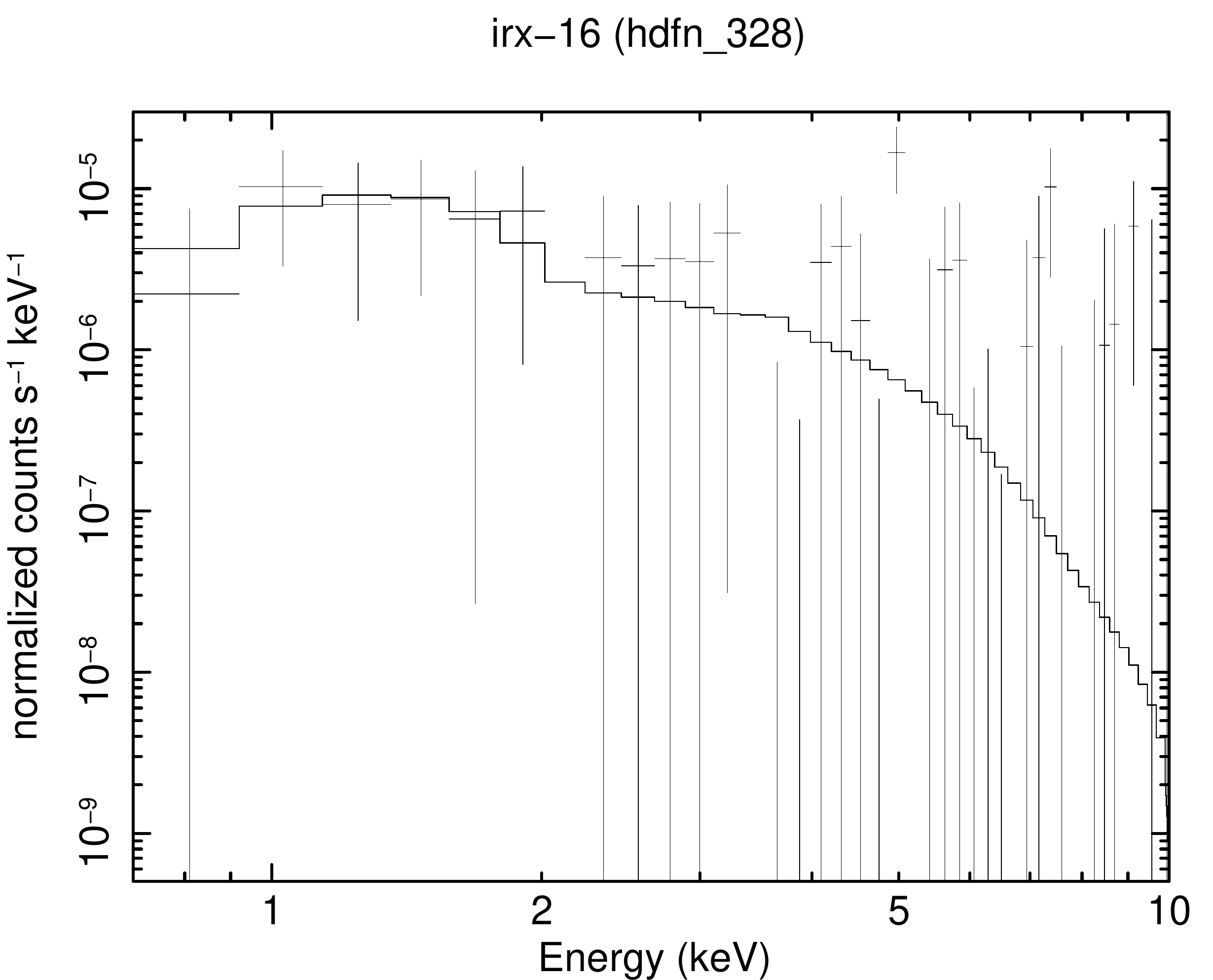}}
\rotatebox{270}{\includegraphics[height=0.8\columnwidth]{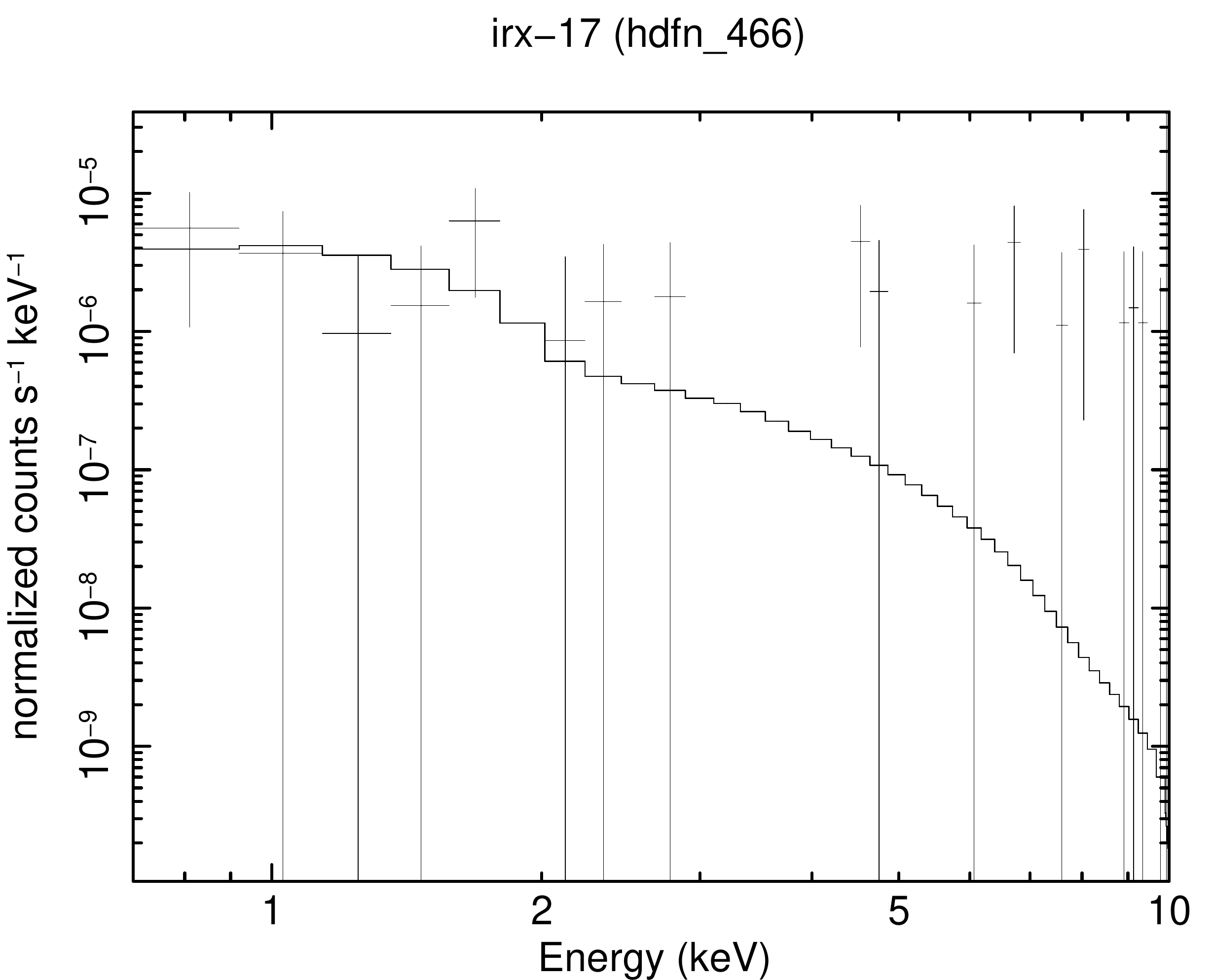}}
\rotatebox{270}{\includegraphics[height=0.8\columnwidth]{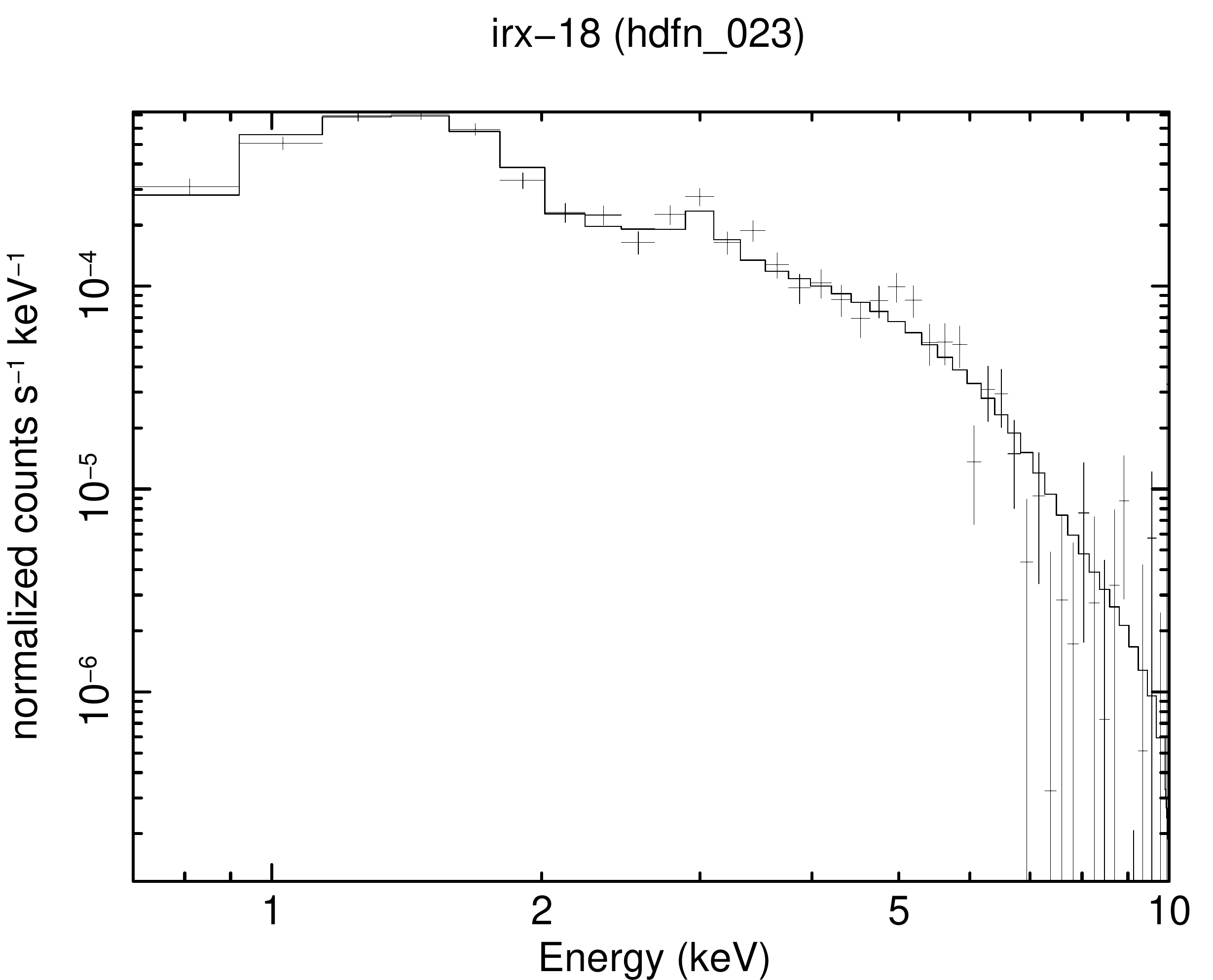}}
\rotatebox{270}{\includegraphics[height=0.8\columnwidth]{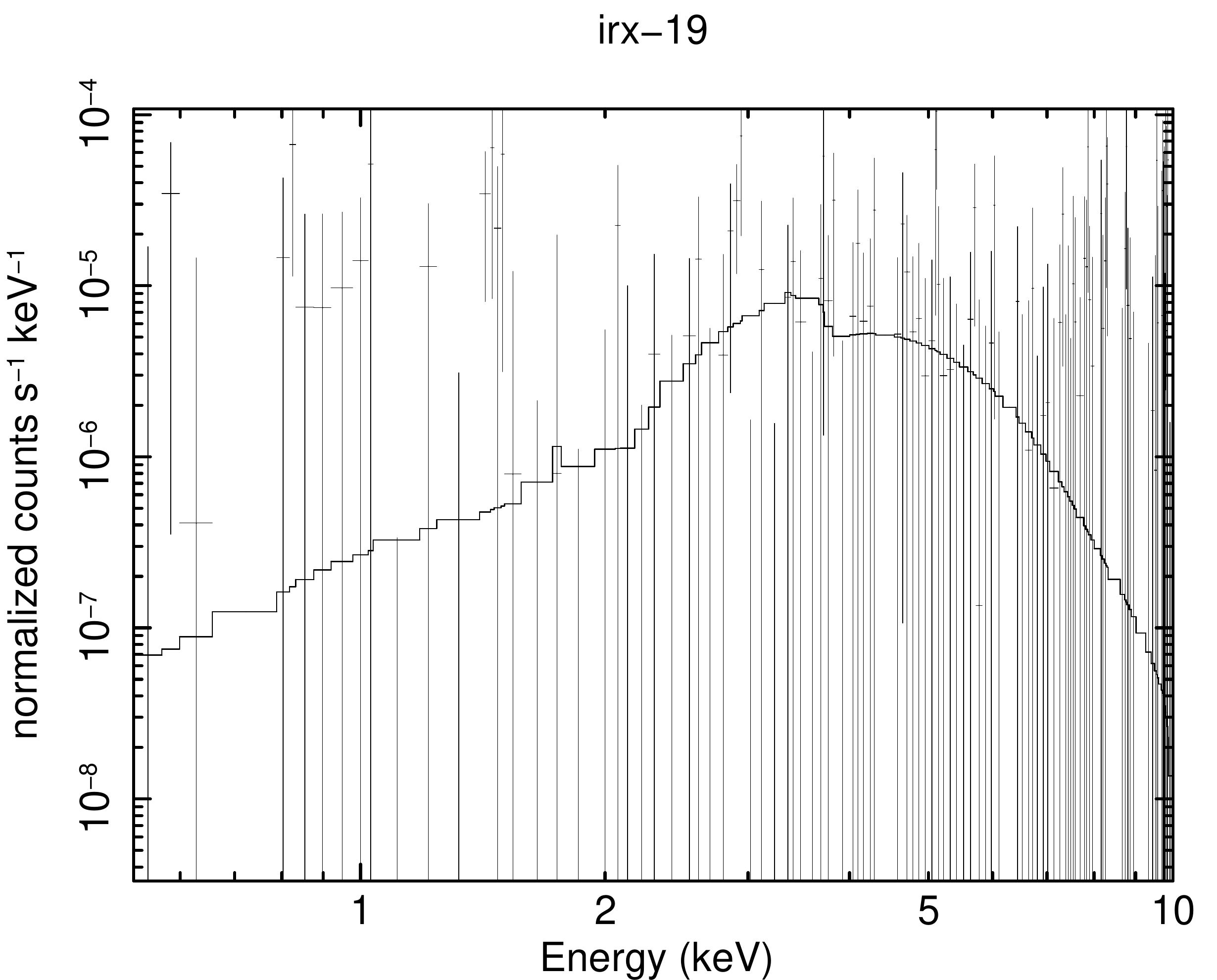}}
\end{center}
\caption{X-ray spectra  and spectral fits  for the IRX sources  in the
  CDF-North.  }\label{fig_xspec2}
\end{figure*}

Spectral analysis was performed  for sources with sufficient number of
net counts to place meaningful constraints.  This requirement excluded
from   the  spectral   fits  sources   with  low   significance  X-ray
counterparts,  except one, irx-19.   This source  lies just  below the
formal  detection threshold  of the  CDF-North source  catalogue.  The
results   of   the  spectral   analysis   are   summarised  in   Table
\ref{tab_xspec}.  Spectral  fits for  individual sources are  shown in
Figures \ref{fig_xspec1}, \ref{fig_xspec2}.  All sources in the sample
are consistent  with moderate column  densities in the range  $\rm N_H
\approx 10^{22} \, - \, 5\times10^{23} \, cm^{-2}$, while the best-fit
relative reflection parameter is $<10$  for all of them. For 4 sources
however, the 90  per cent upper limit in $R$ is  greater than 10.  For
one  of them,  irx-19, the  90  per cent  upper limit  for the  column
density is  $\rm N_H =  1.2 \times 10^{24}  \, cm^{-2}$, close  to the
Compton  Thick  limit ($\rm  N_H  =  2  \times 10^{24}  \,  cm^{-2}$).
Another of the four sources, irx-18, also shows the Fe\,Ka 6.4\,keV in
the X-ray  spectrum with  a rest-frame equivalent  width of $\rm  EW =
0.5\pm0.2$\,keV.   Although a  strong Fe\,Ka  line \citep[$\rm  EW \ga
  1$\,keV;][]{Ghisellini1994} is  evidence for Compton  Thick AGN, the
estimated  EW is  also consistent  with Compton  thin  obscuration. We
conclude that the X-ray spectral analysis suggests that two sources in
the sample show tentative evidence for Compton Thick obscuration.

Also, the  X-ray emission  of two X-ray  detected sources  (irx-16 and
irx-17) is likely associated with starburst activity, not accretion on
a central  supermassive black hole.   These two sources  have $L_X(\rm
2-10\,keV)\approx 5 \times 10^{41} \, erg \, s^{-1}$, relatively soft,
albeit noisy,  X-ray spectra and follow the  correlation between X-ray
and     infrared     luminosities     for    star-forming     galaxies
\citep{Ranalli2003} as  shown in Figure  \ref{fig_ll}.  Their infrared
SEDs are also  consistent with starburst templates and  do not require
an  additional  hot dust  component  associated  with  the AGN  torus.
Figure \ref{fig_ll}  also demonstrates that  a QSO torus  component is
required to  fit the mid-IR part of  the SED of only  those sources in
Table  \ref{tab_sample} which  are detected  at X-ray  wavelengths and
which have X-ray  luminosity, $L_X(\rm 2-10\, keV) \ga  10^{43} \, erg
\,  s^{-1}$.   The SED  fitting  method  of  section \ref{sec_sed}  is
therefore sensitive to AGN  with intrinsic X-ray luminosity above this
limit.

\section{Discussion \& Conclusions}\label{sec_results}

The population  of IRX galaxies has attracted  much attention recently
as  they are proposed  as good  candidates for  Compton thick  QSOs at
$z\approx2$ with luminosities $L_X(\rm 2-10\, keV) > 10^{43} \, erg \,
s^{-1}$.   The low  X-ray--to--mid-IR  luminosity ratio  and the  mean
hardness   ratios  of   these   sources  are   consistent  with   this
interpretation  as  long as  the  bulk  of  the mid-IR  luminosity  is
associated  with the  dust heated  by  the central  engine. There  are
indications however, that  this is not the case, at  least not for all
IRX  sources.   \cite{Murphy2009}  combined mid-IR  spectroscopy  with
far-IR  sub-mm data  to show  that only  50 per  cent of  the infrared
excess sources selected in a way similar to that described by Daddi et
al.  (2007), show evidence for  obscured AGN activity.  The other half
of   this  population  have   SEDs  consistent   with  star-formation.
\cite{Yan2007} presented  Spitzer-IRS spectra of  52 galaxies brighter
than 1\,mJy  at $\rm  24\mu m$ selected  in the  Spitzer extragalactic
First  Look Survey  (xFLS).  The  relative  strengths of  the AGN  and
starburst  components   in  those  sources  have   been  estimated  by
\cite{Sajina2008}  using  in  addition  to  the  mid-IR  spectroscopy,
multiwavelength  photometric observations  (UV to  far-IR  and radio),
optical  and   NIR  spectroscopy.   There   are  21  sources   in  the
\cite{Yan2007} sample with properties similar to the IRX population at
$z\approx2$,  i.e.    $f_{\rm  24\mu  m}/f_R>900$   and  spectroscopic
redshift determination $z>1.5$.  \cite{Sajina2008} found that about 76
per cent (16/25)  of those sources are AGN dominated  in the IR, while
the  remaining  24 per  cent  have  starburst  components that  either
dominate or  contribute nearly equally  to the AGN at  IR wavelengths.
If star-formation contributes to the mid-IR luminosity of IRX galaxies
then  the two  key  properties of  this population,  X-ray--to--mid-IR
luminosity  ratio and mean  hardness ratio,  are also  consistent with
either lower-luminosity Compton thin AGN  (e.g.  Figures 1, 2) or pure
starbursts \citep{Donley2008}.

In order  to shed more  light in the  nature of IRX sources  we select
galaxies at  $z\approx1$ in the  AEGIS and CDF-North fields  with SEDs
similar  to the  $z\approx2$ IRX  population.  The  advantage  of this
approach is that the selected sources have fluxes that are brighter at
almost any wavelength compared to IRX galaxies at $z\approx2$, thereby
greatly facilitating their study.

It is interesting  that about 35 per cent of the  sources in the AEGIS
(5/14) and 80 per cent in the CDF-North (4/5) are associated with hard
(2-7\,keV)  X-ray detections  with significance  $<4  \times 10^{-6}$.
These fractions  increase to 43 (6/14)  and 100 (5/5) per  cent in the
AEGIS  and  CDF-N  respectively,  if lower  significance  sources  are
included. For comparison  only about 4 per cent  of red cloud galaxies
in  the AEGIS  have X-ray  counterparts (Nandra  et al.   2007).  This
suggests a high AGN identification  rate in samples selected using the
infrared excess criteria.

Excluding X-ray detected starburst candidates (irx-16 and irx-17), the
IRX sources  with X-ray counterparts  are indeed AGN, as  indicated by
their  X-ray luminosities,  $L_X(\rm 2-10\,keV)  > 10^{42}  \,  erg \,
s^{-1} \,  cm^{-2}$, which are higher  than what can  be attributed to
star-formation \citep{Georgakakis2007}.  These  sources also have hard
X-ray spectral  properties, which are however  consistent with Compton
Thin column densities,  $N_H \approx \rm 10^{22} -  5\times 10^{23} \,
cm^{-2}$.   There  is  only   tentative  evidence  for  Compton  Thick
obscuration among  the X-ray detected  IRX sources.  For four  of them
the 90 per cent upper limit  of the reflection fraction is higher than
$R=10$.  Also, the  X-ray spectra of two of  those four sources either
have   90    per   cent   upper   limit   in    the   column   density
$\approx1.2\times10^{24}    \rm    \,    cm^{-2}$    or    show    the
FeK$\alpha$\,6.4keV line with an equivalent width of $0.5$\,keV. These
properties can  be interpreted as  evidence for Compton Thick  AGN but
are also consistent with Compton thin obscuration.

More relevant to the deeply buried AGN picture are the infrared excess
sources in the sample that are not detected at X-ray wavelengths.  The
SED  modelling  shows that  their  mid-  and  far-IR is  dominated  by
star-formation  and  that  there  is  no indication  for  a  hot  dust
component associated  with a powerful  AGN that remains  undetected at
X-ray wavelengths.  It is noted  the QSO torus template is required to
fit the IR  SEDs of {\it all} X-ray detected  AGN with intrinsic X-ray
luminosities brighter than $L_X(\rm  2-10\, keV)\approx 10^{43} \, erg
\, s^{-1}$.  Therefore,  if some of the X-ray  undetected sources were
associated with heavily obscured  and powerful AGN with X-ray emission
suppressed  by the  intervening dust  and  gas clouds,  we would  have
identified them in  the mid-IR as sources with  a QSO torus component.
We  can therefore,  place an  upper limit  of $L_X(\rm  2-10\,  keV) =
10^{43} \,  erg \,  s^{-1}$ to the  intrinsic AGN luminosity  of X-ray
undetected sources, if they host  an active SBH.  We conclude that the
IRX  sources  without X-ray  counterparts  in  our  sample are  either
starbursts, lower luminosity AGN, or a combination of the two.

This  result has  implications on  the nature  of the  $z\approx2$ IRX
population detected  in deep surveys. There  is no doubt  that some of
these sources  are Compton Thick  QSOs. \cite{Georgantopoulos2009} for
example,  identified Compton  Thick QSOs  among X-ray  sources  in the
CDF-North  through X-ray  spectroscopy and  showed that  some  of them
satisfy the infrared excess selection criteria.  Our analysis however,
shows  that a  potentially large  fraction  of the  IRX population  at
$z\approx2$ are not luminous [$L_X(\rm 2-10\, keV) > 10^{43} \, erg \,
  s^{-1}$] Compton  Thick QSOs  but lower luminosity  [$L_X(\rm 2-10\,
  keV) < 10^{43}  \, erg \, s^{-1}$] possibly  Compton thin AGN and/or
starbursts.

Dust  enshrouded  star-formation  has  already  been  proposed  as  an
alternative to  Compton Thick  QSOs to explain  the properties  of the
$z\approx2$   IRX  sources.   \cite{Donley2008}   critically  reviewed
different  methods  proposed in  the  literature  for  finding AGN  at
infrared wavelengths,  including the infrared  excess selection.  They
showed that the red SED of  these sources are consistent with those of
local  pure  starbursts  with  some  moderate  amounts  of  additional
reddening  at   optical  wavelengths  ($A_V\sim1$\,mag).    They  also
cautioned  that the  stacked X-ray  signal of  this population  may be
dominated by few  sources. Moreover, they argued that  flat mean X-ray
spectral properties of the IRX  population could be the result of high
mass  X-ray binaries,  which are  known  to have  hard X-ray  spectral
properties  (power-laws with  $\Gamma \approx  1.2$).   Whether binary
stars can  dominate the integrated X-ray emission  of local starbursts
however, is still  under debate.  In any case,  based on the arguments
above  \cite{Donley2008} suggested  that as  much as  half of  the IRX
population at $z\approx2$  are starbursts with only about  20 per cent
showing  evidence for  heavily obscured,  possibly Compton  Thick, AGN
activity.   \cite{Pope2008}  analysed the  mid-IR  spectra  of 12  IRX
sources with $f_{\rm 24\mu m}>300\, \rm \mu Jy$ in the GOODS-North and
found that  6 of them are  dominated by star-formation  at the mid-IR.
They also  showed that  AGN and starburst  dominated IRX  sources have
distinct  $\rm 8.0\mu  m$ over  $\rm  4.5\mu m$  flux ratios,  $f_{\rm
  8.0\mu m}/f_{\rm 4.5\mu m}$.  Extrapolating their results to fainter
IRX sources $f_{\rm  24\mu m}=100-300\, \rm \mu Jy$,  for which mid-IR
spectroscopy is  not available, they found  that about 80  per cent of
them  have  $f_{\rm  8.0\mu   m}/f_{\rm  4.5\mu  m}$  consistent  with
star-formation.  They  also estimate  the mean SED  of IRX  sources in
their sample  and conclude  that less  than about 10  per cent  of the
total infrared  luminosity ($\rm 8-1000\mu m$) is  associated with hot
dust, possibly heated by an  AGN.  For the average infrared luminosity
of  the  \cite{Pope2008} sample,  $L_{IR}=10^{12}  \, L_{\odot}$,  the
fraction above translates  to an upper limit in  the AGN luminosity of
$L_{IR}\approx4\times 10^{44} \rm \, erg \, s^{-1}$.  Adopting the AGN
bolometric      correction       factors      $L_{bol}/L_{IR}=      3$
\citep{Risaliti_Elvis2004} and $L_{bol}/L_{X}(\rm 2  - 10 \, keV)= 35$
\citep{Elvis1994},  we estimate  a mean  hard X-ray  luminosity $L_{X}
(\rm 2-10 \,keV) < 3\times  10^{43} \, erg \, s^{-1}$.  Although there
are uncertainties in this calculation, there is broad agreement in the
upper limits in $L_X$ estimated for the \cite{Pope2008} and our sample
of IRX sources.

The sample presented here  has properties similar to moderately bright
IRX sources  at $z\approx2$,  i.e.  $\rm 5.8\,  \mu m$  luminosity few
times $\rm 10^{44} \, erg \,s^{-1}$, like those found in the CDF-South
by  Fiore et  al.   (2008).  The  simulations of  \cite{Narayanan2009}
suggest  that IRX  sources above  $S_{24}  \rm =  300 \,  \mu Jy$  are
dominated by gaseous mergers and  include a large fraction of powerful
AGN,  whereas less  luminous  systems, like  those  studied here,  are
typically secularly evolving  galaxies dominated by star-formation and
with only weak  AGN activity. Our results on  the nature of moderately
luminous IRX  sources are therefore consistent  with those simulation.
It is  likely that IRX sources  more luminous than  those studied here
(i.e. COSMOS  field, $S_{24}  \rm > 500  \, \mu Jy$,  $\nu L_{\nu}(\rm
5.8\mu m)>  10^{45} \,  erg \,s^{-1}$; Fiore  et al.  2009)  include a
higher  fraction of  Compton Thick  QSOs.   The results  of Sajina  et
al. (2008) on the nature of bright  ($S_{24} \rm > 900 \, \mu Jy$) IRX
sources selected in the xFLS (Yan et al.  2007) suggests that this may
be  the case.   \cite{Bauer2010} explored  the X-ray  properties  of a
subset of  the Yan  et al.   (2007) sample and  found evidence  for at
least  mildly   Compton  Thick  ($N_H\approx10^{24}\,   \rm  cm^{-2}$)
obscuration in a large fraction  of the sources that are AGN dominated
in the  mid-IR. Even  at the extreme  luminosities of the  xFLS sample
however, a  non-negligible fraction of  the IRX sources (about  24 per
cent) have a  substantial or even dominant starburst  component in the
mid-IR.    This  underlines   the  importance   for   subtracting  the
contribution of  star-formation to the mid-IR to  assess the intrinsic
AGN luminosity  before concluding whether  the X-ray/mid-IR properties
of these systems are consistent with Compton Thick obscuration.

\cite{Mullaney2009}  have  recently  found  that the  X-ray  to  total
infrared luminosity ratios of X-ray AGN with $L_{X} (\rm 2-10 \,keV) =
10^{42} -  10^{43} \,  erg \, s^{-1}$  increases by about  1\,dex from
$z=0$ to $z\approx1-2$.  Although the origin of this trend is unclear,
it can be interpreted as  an enhancement of the average star-formation
rate  in lower-luminosity  AGN at  $z\approx2$ compared  to  the local
Universe.  In  this picture, the enhanced star-formation  is likely to
have  an impact on  the mid-IR  part of  the SED  and will  make lower
luminosity  AGN at  $z\approx1-2$ appear  underluminous at  X-rays for
their mid-IR luminosity compared to  local AGN samples.  This would be
in  agreement  with  our  interpretation  of the  IRX  population  and
emphasises the  need to properly model  the mid-IR part of  the SED to
assess  the relative contribution  of AGN  and star-formation  at this
wavelength regime.

Further  progress  in the  study  of  IRX  galaxies is  expected  from
upcoming Herschel observations, which  when combined with the existing
Spitzer  data  will  constrain  the  SEDs  of  individual  sources  at
$z\approx2$  over a  sufficiently  large wavelength  baseline (mid  to
far-IR) to  allow decomposition of  the AGN and  starbursts components
through  template fits,  as done  in this  paper.  This  exercise will
provide estimates  of the level  of dusty star-formation  activity and
the  intrinsic AGN  luminosity  of individual  sources  to confirm  or
refute claims that they are  Compton Thick QSOs.  The determination of
the column density of these  sources through X-ray spectroscopy has to
wait  future X-ray  missions with  large collecting  areas,  such IXO,
which is expected to provide spectra for $z\approx2$ IRX sources which
have typical X-ray fluxes $f_X ( \rm  2 - 10 \, keV ) \approx 10^{-17}
\,  erg  \, s^{-1}  \,  cm^{-2}$  \citep{Georgantopoulos2009} and  are
currently accessible only through stacking analysis.

\section{Acknowledgments} The  authors wish  to thank the anonymous
referee for constructive suggestions that improved the paper and  Alison  Coil for
useful  comments.   AG   acknowledges  financial  support  from  the
Marie-Curie  Reintegration Grant  PERG03-GA-2008-230644.   P.G.  P.-G.
acknowledges  support  from  the   Spanish  Ministry  of  Science  and
Innovation  under grants  AYA 2006-02358,  AYA  2006-15698-C02-02, and
CSD2006-100070, and from  the Ram\'on y Cajal Program  financed by the
Spanish Government  and the  European Union. This  study makes  use of
data  from AEGIS,  a  multiwavelength sky  survey  conducted with  the
Chandra,  GALEX, Hubble,  Keck, CFHT,  MMT, Subaru,  Palomar, Spitzer,
VLA, and other telescopes and supported  in part by the NSF, NASA, and
the STFC.

\bibliography{mybib}{}
\bibliographystyle{mn2e}

\end{document}